\pdfoutput=1 
\documentclass[singlespacing]{elsart}
\usepackage{graphics}
\usepackage{graphicx}
\usepackage{epsfig}
\usepackage{amssymb}
\begin{document}
\begin{frontmatter}
\title{An Immersed Boundary Fourier Pseudo-spectral Method for Simulation of Confined Two-dimensional Incompressible Flows}
\author{Fereidoun Sabetghadam\corauthref{cor1}}
\ead{fsabet@srbiau.ac.ir}
\corauth[cor1]{}
\author{Mehdi Badri, Shervin Sharafatmandjoor, Hosnieh Kor}
\address{Mechanical and Aerospace Engineering Department, Science and Research Branch, Azad University (IAU), Tehran, Iran.}
\begin{abstract}
The present paper is devoted to implementation of the immersed boundary technique into the Fourier pseudo-spectral solution of the vorticity-velocity formulation of the two-dimensional incompressible Navier--Stokes equations. The immersed boundary conditions are implemented via direct modification of the convection and diffusion terms, and therefore, in contrast to many other similar methods, there is not an explicit external forcing function in the present formulation. The desired immersed boundary conditions are approximated on some regular grid points, using different orders (up to second-order) polynomial extrapolations. At the beginning of each timestep, the solenoidal velocities (also satisfying the desired immersed boundary conditions), are obtained and fed into a conventional pseudo-spectral solver, together with a modified vorticity. The zero-mean pseudo-spectral solution is employed, and therefore, the method is applicable to the confined flows with zero mean velocity and vorticity, and without mean vorticity dynamics. In comparison to the classical Fourier pseudo-spectral solution, the method needs ${\cal O}(4(1+\log N)N)$ more operations for boundary condition settings. Therefore, the computational cost of the method, as a whole, is scaled by ${(N \log N)}$. The classical explicit fourth-order Runge--Kutta method is used for time integration, and the boundary conditions are set at the beginning of each sub-step, in order to increasing the time accuracy. The method is applied to some fixed and moving boundary problems, with different orders of boundary conditions; and in this way, the accuracy and performance of the method are investigated and compared with the classical Fourier pseudo-spectral solutions.
\end{abstract}
\begin{keyword}
Fourier pseudo-spectral solution; Two-dimensional Navier--Stokes equations; Vorticity--velocity formulation;
 Immersed boundary method; Solenoidal velocities; Moving boundary problems  \end{keyword}

\end{frontmatter}
\section{Introduction}
\label{s1}
Fourier pseudo-spectral solution of the vorticity-based formulations of the Navier--Stokes equations (NSE) have been used~ widely in the~ two-dimensional incompressible flow simulations \cite{Canuto1}. However, the classical implementations are limited to the regular domains with simple coordinate-coinciding boundaries and periodic boundary conditions. Now, recent advances in the immersed and embedded boundary techniques have raised hopes of extending the range of applicability of these methods to the more general domains and boundary conditions \cite{Boyd05,Sabetghadam,Kolomenskiy}.\\
With the best knowledge of the authors, the immersed boundary method was applied into the vorticity--stream function formulation of the NSE by Calhoun \cite{Calhoun1,Calhoun2} for the first time. In the Calhoun's work the immersed surfaces are introduced, and the overall mass balance is satisfied, by definition of an appropriate distribution of vorticity source term. More or less similar line was followed in the work of Russell and Wang \cite{Russel}. They decomposed the effects of a solid wall into a no-slip condition, satisfied by the Thom's rule, and a no-penetration condition, imposed by a boundary element method (which satisfied the overall mass balance). In the work of Linnick and Fasel \cite{Fasel}, a higher-order compact method was used, and a source term was defined in crossing the discontinuities, which was obtained from a jump function. Recently, Wang {\it et al.} \cite{Wang} applied the {\it direct forcing} idea of Mohd-Yusof \cite{Mohd} into the vorticity--velocity formulation of the NSE. They added an explicit vorticity source term to the vorticity transport equation, which was obtained by taking curl of the forcing functions of the momentum equations in the primitive variables form of the NSE. However, all the above methods were  based upon the finite-difference or finite-volume spatial discretization.\\
In the pseudo-spectral solutions, the volume penalization is one of the popular remedies, which has been used several times for implementation of the no-slip condition. The method was first proposed in the primitive variables formulation of the NSE by Arquis and Caltagirone \cite{Arquis}, and then re-formulated by Angot \cite{Angot}. In the next years the method was extended to the vorticity--velocity formulation \cite{Kevlahan,Schneider1,Keetels}, and used for the fixed, as well as moving boundary problems \cite{Kolomenskiy,Schneider2,Kevlahan,Clercx97}.\\
A new immersed boundary method is proposed in the present paper, in which the arbitrary immersed velocity boundary conditions (including the no-slip condition), are introduced into the Fourier pseudo-spectral solution of the vorticity--velocity formulation of the NSE, without explicit addition of external forcing functions. Instead of the conventional forcing functions, the immersed boundaries are implemented by direct modification of the convection and diffusion terms of the vorticity transport equation in such a way that can be implemented to the Fourier pseudo-spectral solutions.\\
\begin{figure}[t]
\setlength{\unitlength}{1mm}
\centerline{\includegraphics[width=8.5cm]{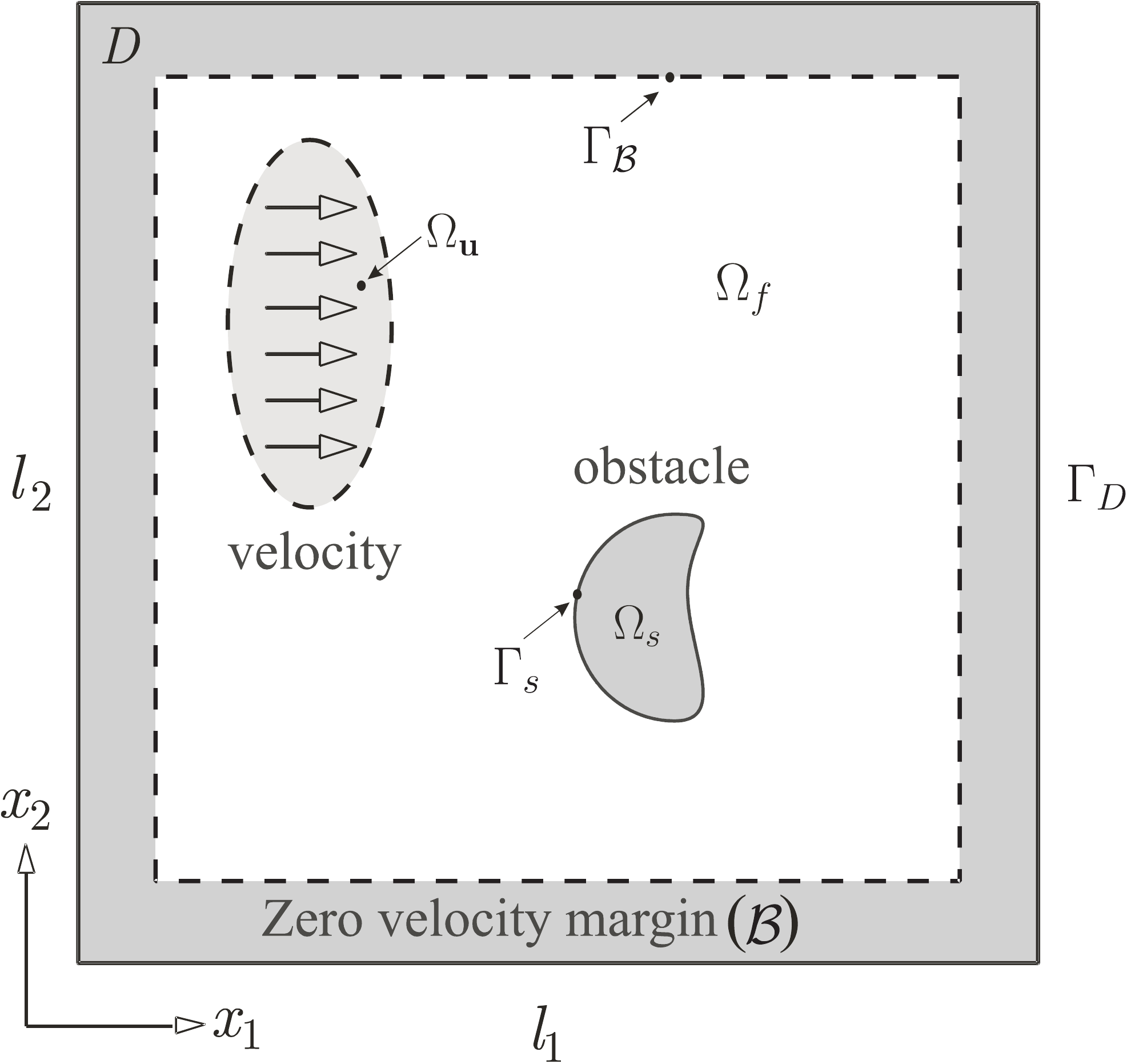}}
\caption{The flow domain $\Omega_f$, together with fixed or moving obstacle(s) $\Omega_s$, and other given-velocity regions $\Omega_{\rm{\bf u}}$, are embedded in the regular solution domain $D$ (with regular boundary  $\Gamma_{D}$), via a zero-velocity margin $\cal B$.} \label{fig1} \vspace{7mm}
\end{figure}
Particularly, it is aimed to use the zero-mean Fourier pseudo-spectral method. Therefore, among many possibilities, our suggested flow configuration is illustrated in Fig.~\ref{fig1}. The regular solution domain $D$ contains all the fluid--solid system, including the moving regions of the fluid with given velocity boundary conditions (that is $\Omega_{\rm{\bf u}}$), in addition to the fixed or moving immersed solid objects. Moreover, note that the fluid--solid system is embedded in $D$ via a zero-velocity margin $\cal{B}$. This margin is considered to ensure that the flow physics can be followed by the zero-mean Fourier series, namely, to ensure that the velocity and vorticity fields remain zero-mean, and the dynamics of the mean vorticity is zero, during the time integration. In fact, we found that considering such a margin can also be useful in other methods ({\it e.g.} the finite difference method), in order to overcoming some difficulties related to the appropriate vorticity boundary conditions and finding solenoidal velocities (see the appendix for more discussions about the role of margin $\cal B$). In the present work, this margin is obtained by windowing of the velocity fields.\\
On the other hand, definition of this margin is also in the following of our previous work \cite{Sabetghadam}, and the idea of an {\it infinitely flat} window function, proposed in \cite{Boyd05}. However, achieving the spectral rates of convergence is not anticipated for the finite Reynolds numbers (which is the subject of this paper); therefore, it is not needed {\it infinitely} flat window functions. However, the algebraic rates of convergence of the Fourier series are competing in many practical situations. Although the shape of the inner boundary of $\cal B$ ({\it i.e.} $\Gamma_{\cal B}$) is arbitrary, in our numerical experiments we used rounded rectangular shapes for simplicity.\\
The central core of the method is a conventional pseudo-spectral solver of the two-dimensional incompressible NSE in the vorticity-velocity form. At the beginning if each time step, the desired immersed velocity boundary conditions are imposed via direct modification of the velocity fields. Then the modified vorticity (which will be called the conditioned vorticity), is constructed directly from the modified velocities. The modified vorticity is fed directly into the pseudo-spectral solver, which gives the vorticity field in the next time.\\
Time integration is performed using the classical explicit fourth-order Runge--Kutta method to keep the method as simple as possible, and to avoid some difficulties associated with the implicit formulations. In comparison to the classical pseudo-spectral method, the present method needs ${\cal O} (4(1+\rm log N)N)$ more operations for boundary condition settings, and therefore, the computational cost of the method, as a whole, is scaled by ${(\rm N \log N)}$.
\\
The paper is continued by presenting the mathematical formulations of the classical Fourier pseudo-spectral method and the suggested algorithm for imposing the immersed velocity boundary conditions. Because of its crucial role, the boundary condition setting process is presented in details, in an individual section. As our numerical experiments, the method is implemented into some fixed as well as moving boundary problems, with the surfaces which are coinciding and non-coinciding with the regular grids. Finally, some basic questions about the validity of the method are addressed in an appendix.\\
\section{Mathematical Formulation}
\label{s2}
The mathematical and numerical frameworks of the method are described in this section. Beginning from the classical Fourier pseudo-spectral formulation, the suggested modifications for imposing the immersed boundaries, and then embedding the solution domain into the regular domain are explained.
\subsection{The Fourier pseudo-spectral formulation} \label{sec2.1}
According to Fig. \ref{fig1}, for a two-dimensional velocity vector ${\bf u}=(u_1,u_2)$, defined on the regular closure  $\bar{D}=(D \cup \Gamma_D)$, the dynamics of the vorticity vector ${\bf \omega}=(0,0,\omega_z=\omega {\hat {\bf e}}_{z}=\nabla\times{\bf u})$ is obtainable from time integration of the vorticity transport equation
\begin{equation}
\cases{
\partial_t \omega+({\bf u}\cdot \nabla)\omega=\nu\nabla^2\omega \quad \quad {\rm in}\quad D\times (0,T], \label{e1} \cr
\omega ({\bf x},t=0)=\omega_0 ({\bf x}) \quad \quad \quad ~~ {\rm for}\quad {\bf x}\in \bar{D}, \cr }
\end{equation}
while the velocity vector $\bf u$ satisfies the following Poisson's problem with Dirichlet boundary conditions:
\begin{equation}
\cases{
\nabla^2 {\bf u} ={\hat {\bf e}}_{z}\times\nabla\omega \quad \quad {\rm in}\quad D, \label{e2} \cr
{\bf u}(\Gamma_D)={\bf u}_{\Gamma_D}.\cr}
\end{equation}
One of the advantages of the above vorticity-velocity formulation, in comparison to {\it e.g.} many primitive variable formulations, is the possibility of decomposition of the kinematics and dynamics of the flow field at each time instant. In fact, for any arbitrary distribution $\omega\in{\cal L}^2(D)$, the physical (divergence-free) velocity vector is obtainable from solution of Eq. (\ref{e2}), if the appropriate boundary conditions are imposed (see the appendix or \cite{Dennis} for more discussions). As it will be seen, this issue has a vital role in construction of the physical immersed velocities in the present method.\\
On the other hand, to improve the efficiency of the computations, it is convenient to change the vorticity transport equation (\ref{e1}) to
\begin{equation}
\partial_t \omega=\underbrace{\nu\nabla^2\omega}_{\rm{\bf L}} -\underbrace{\frac{\partial^2}{\partial x_1\partial x_2}(u_2^2-u_1^2)}_{{\rm{\bf N}}_1}+\underbrace{(\frac{\partial^2}{\partial x_2^2}-\frac{\partial^2}{\partial x_1^2})u_1u_2}_{{\rm{\bf N}}_2},\label{e1mod}
\end{equation}
which in comparison to the classical form (\ref{e1}), saves one fast Fourier transform (FFT) in the pseudo-spectral algorithm. Although this formulation has been used in some other studies, to the best knowledge of the authors, in the pseudo-spectral solution of the incompressible flow, it was proposed by Chasnov \cite{Chasnov} for the first time. The diffusion term ${\rm{\bf L}}$ and the non-linear terms, ${\rm{\bf N}}_1$ and ${\rm{\bf N}}_2$, are named for the future references. In fact, the immersed velocity boundary conditions will be introduced to the solution by direct modification of these terms.\\
For the periodic boundary conditions, the Fourier series provides such an accurate and efficient tool which makes it worthwhile  re-formulating the problem in the Fourier space. In this way, the vorticity transport equation (\ref{e1mod}) recasts
\begin{equation}
\cases{
d_t \hat{\omega}=-\underbrace{\nu |{\bf k}|^2\hat{\omega}}_{\hat{\rm{\bf L}}}-\underbrace{k_1k_2\widehat{(u_1^2-u_2^2)}}_{\hat{{\rm{\bf N}}}_1}+\underbrace{(k_1^2-k_2^2)\widehat{u_1u_2}}_{\hat{{\rm{\bf N}}}_2}, \cr
\hat{\omega}({\bf k},t=0)=\hat{\omega}_0, \cr} \label{e3}
\end{equation}
while the Poisson's problem (\ref{e2}) simplifies to
\begin{equation}
\hat{\bf u}=-i\frac{{\bf k}^{\bot}}{|{\bf k}|^2}\hat{\omega}. \label{e5}
\end{equation}
In these equations, $\hat{(\cdot)}$ stands for the quantities in the Fourier space, ${\bf k}=(k_1,k_2)$ is the wavenumber vector with the magnitude $|{\bf k}|^2=k_1^2+k_2^2$; while ${\bf k}^{\bot}=(-k_2,k_1)$ is the perpendicular wavenumber vector, and $i^2=-1$. Practically, in the finite dimensional calculations, the nonlinear terms ${\hat{{\rm{\bf N}}}}_1$ and ${\hat{{\rm{\bf N}}}}_2$, are constructed (and are de-aliased) in the physical space---  the algorithm which is known (and will be referred in this paper) as the pseudo-spectral method. Discretization in time, and time integration of the fully discretized system can be done using an appropriate time marching method. The explicit fourth-order Runge--Kutta method is used in the present work.\\
More or less similar formulations have been used in many efficient and accurate solvers of the periodic flow in regular domains. In the sequel we will modify equations (\ref{e3}) and (\ref{e5}), and suggest an algorithm to use them in the flow configuration of Fig. \ref{fig1}.
\subsection{Implementation of the immersed velocity boundary conditions}\label{sec2.3}
In the present method, without explicit addition of a forcing function in the right hand side of Eq. (\ref{e3}), the immersed surfaces are introduced by modification of the $\hat{{{\rm{\bf N}}}}_1$, $\hat{{{\rm{\bf N}}}}_2$, and $\hat{\rm{\bf L}}$. Particularly, it is desired to carry out these modifications such that the velocities remain solenoidal, and in a manner that the method can be implemented easily into a Fourier pseudo-spectral solver. The suggested procedure is summarized in Fig. \ref{fig4}, in which the boundary conditions box (BC) can be more explained by the following remarks:
\begin{figure}[t]
\setlength{\unitlength}{1mm}
\centerline{\includegraphics[width=11cm]{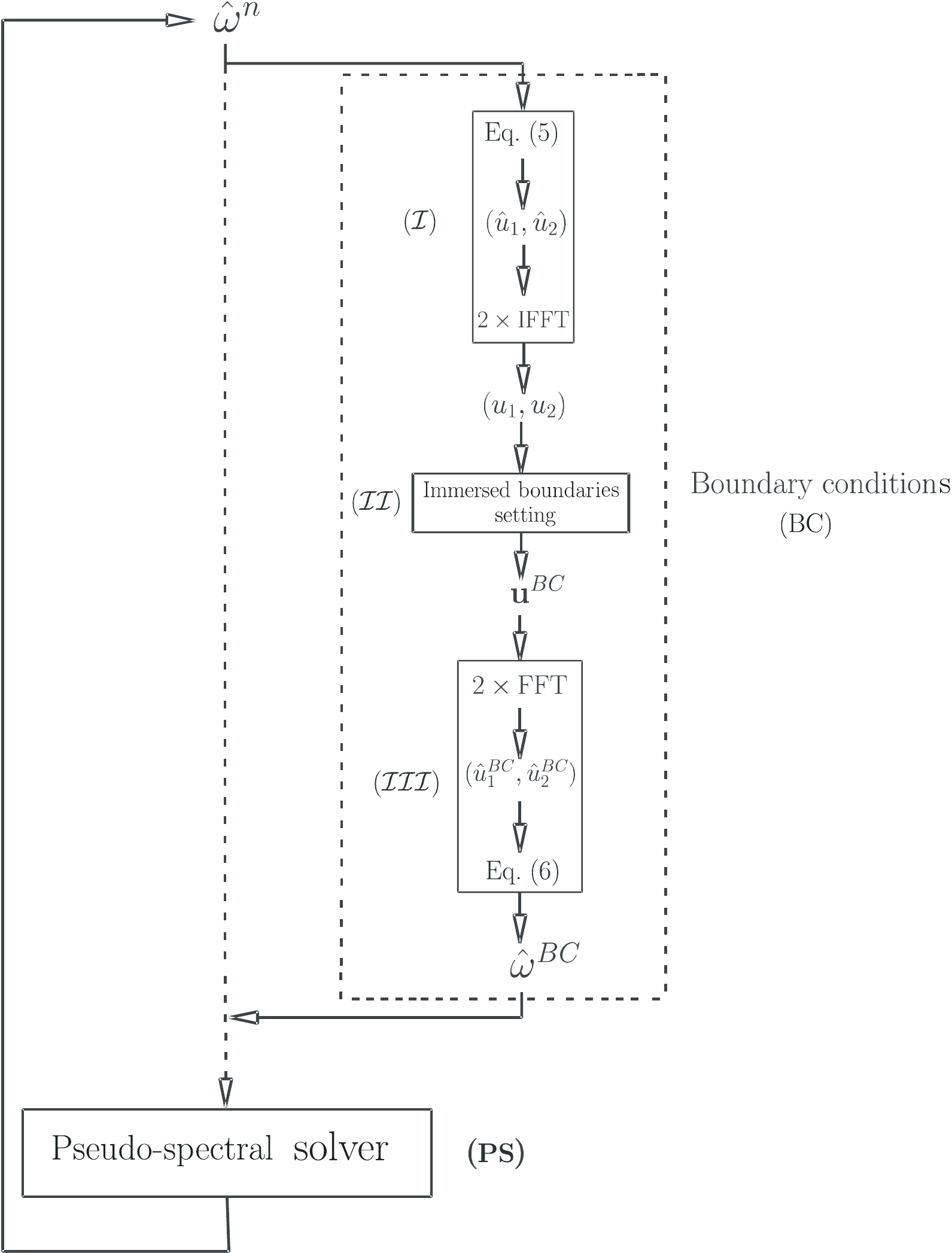}}
\caption{The main steps of the proposed algorithm. The box {\bf (PS)} contains a classical Fourier pseudo-spectral solver (calculation of the right hand side of Eq. (\ref{e3}), solution of Eq. (\ref{e5}), de-aliasing, time integration\ldots). Therefore, a classical Fourier pseudo-spectral solution with periodic boundary conditions can be retrieved by switching-off the boundary condition setting box, and following the dashed line.} \label{fig4} \vspace{5mm}
\end{figure}
\begin{itemize}
\item[1.] Given the vorticity field $\hat{\omega}^n$, from the initial condition or the last timestep, the velocity vector $\bf{u}$ in the regular domain $\bar D$, is obtained from Eq. (\ref{e5}) and two inverse FFTs (box {$\cal (I)$}).
\item[2.] The velocity vector $\bf{u}$ is modified to satisfy all needed immersed velocity boundary conditions (box $\cal (II)$). This conditioned velocity will be called ${\bf u}^{\rm BC}$.\\
The modifications are carried out by local extrapolations, extension, and windowing of $\bf{u}$; and will be explained in details in $\S$ \ref{2.3}. Note that ${\bf u}^{\rm BC}$ is neither necessarily solenoidal, nor its mean value is necessarily zero at this point.
\item[3.] The conditioned vorticity $\omega^{\rm BC}$ is re-calculated from ${\bf u}^{\rm BC}$ (that is,  $\omega^{\rm BC}=\nabla\times{\bf u}^{\rm BC}$), as it is shown in box $\cal (III)$.\\
There are two main reasons for this step. Firstly, the solenoidal velocities can be obtained from this conditioned vorticity in the next steps, provided that the appropriate boundary conditions are implemented (see the appendix and the discussions in there); and secondly, the vorticity will be needed in the subsequent pseudo-spectral algorithm.\\
Although ${\omega}^{\rm BC}$ can be found by any method ({\it e.g.} the finite difference), to preserve the spectral accuracy, and because the vorticity in the Fourier space is needed in the subsequent steps of the pseudo-spectral algorithm, calculation in the Fourier space is suggested here. In this way
\begin{equation}
\hat{\omega}^{\rm BC}=i(k_1\hat{u}_2^{\rm BC}-k_2\hat{u}_1^{\rm BC}). \label{e7}
\end{equation}
Note that it is aimed to simulate the confined flows, and therefore, the vorticity field has zero-mean according to the Stokes theorem; the fact that legitimates use of the above equation. Moreover, note that $\omega^{\rm BC}={\rm IFFT}\{\hat{\omega}^{\rm BC}\}$ is automatically defined on ${\bar D}$, it is double periodic, and it has zero mean--- the properties that makes it {\it ready to use} for the subsequent Fourier pseudo-spectral steps.
\item[4.] The conditioned vorticity $\hat{\omega}^{\rm BC}$ is fed into the classical pseudo-spectral procedure (that is, box {\bf (PS)} in Fig. \ref{fig4}), in which the solenoidal velocity vector $\hat{\bf u}_{\bf Sol}^{\rm BC}$ is calculated from
\begin{equation}
\hat{\bf u}_{\bf Sol}^{\rm BC}=-i\frac{{\bf k}^{\bot}}{|{\bf k}|^2}\hat{\omega}^{\rm BC}; \label{e5p}
\end{equation}
and these velocities, in addition to the conditioned vorticity $\hat{\omega}^{\rm BC}$ are substituted into the modified vorticity transport equation
\begin{equation}
\cases{
d_t \hat{\omega}=-\nu |{\bf k}|^2\hat{\omega}^{\rm BC}-k_1k_2\widehat{(u_1^2-u_2^2)}_{\bf Sol}^{\rm BC}+(k_1^2-k_2^2)\widehat{(u_1u_2)}_{\bf Sol}^{\rm BC}, \cr
\hat{\omega}({\bf k},t=0)=\hat{\omega}^{\rm BC}. \cr} \label{e3p}
\end{equation}
Time integration of the above equation yields the new vorticity field $\hat{\omega}^{n+1}$ which closes the algorithm loop.
\end{itemize}
An attractive feature of the above procedure is that the boundary conditions box can be added easily to a classical pseudo-spectral solver without any change in the internal structure. In fact, the box ({\bf PS}) contains all steps of a classical pseudo-spectral solution (as it was formulated in $\S \ref{sec2.1}$). Moreover, note that in addition to the solid obstacles, the given immersed fluid velocities can be implemented as well.\\
On the other hand, as it was mentioned earlier, implementation of the boundary conditions (that is, box $\cal (II)$), can imposes some discontinuities to the velocity field. Although the next steps will remove these discontinuities (as it will be proven in the appendix), the boundary conditions will change a bit. Similar problem was observed in many other immersed boundary methods, and caused emergence of some methods like the multi-direct forcing method \cite{Wang, Luo}, or the implicit forcing method \cite{Le, Uhlmann} in the last years. In the present work, we repeat the above algorithm  ${\cal N}_r$ times to overcome this difficulty; where $1 \leq{\cal N}_r\leq 3$ depending on the flow. By ${\cal N}_r$ times repetition of the boundary condition settings (and also by development of the solution), the velocity field converges to a solenoidal field which satisfies (approximately) the desired velocity boundary conditions.\\
In our numerical experiments that are presented in this paper, the fourth-order Runge--Kutta method is used for time integration, and in order to increasing the time accuracy, the boundary conditions are set at the beginning of each Runge--Kutta sub-step.
\begin{figure}[t]
\setlength{\unitlength}{1mm}
\centerline{\includegraphics[width=8cm]{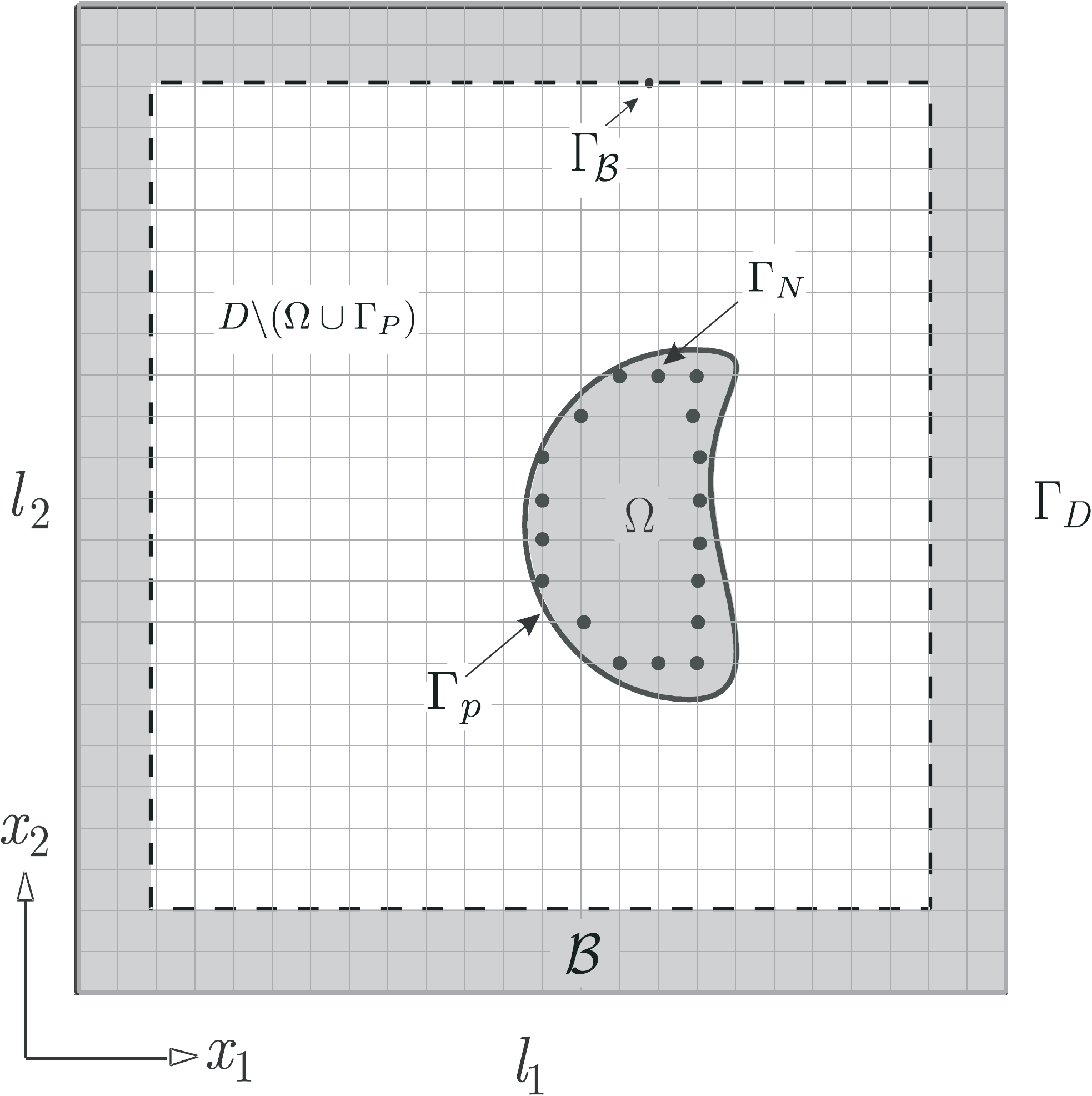}}
\caption{The numerical boundary $\Gamma_{\rm N}$ corresponding to the physical boundary $\Gamma_{\rm P}$. The immersed boundary conditions ${\bf u}_{\rm pb}$ are given on $\Gamma_{\rm P}$.} \label{fig3}
\vspace{5mm}
\end{figure}
\subsection{Boundary conditions setting}\label{2.3}
This section is devoted to a full description of the method used in setting of the immersed velocity boundary conditions (that is, modifying $\bf u$ into ${\bf u}^{\rm BC}$, mentioned in item 2 of section \ref{sec2.3}). The method includes a local extrapolation, borrowed from the finite difference-based immersed boundary methods \cite{Sjogreen}, followed by an extension and windowing process, which we used in our previous work \cite{Sabetghadam}. Particularly, this combination is chosen in order to achieving different (desired) orders of {\it local accuracies} in boundary condition settings, and the needed flexibility in treating the moving boundary problems. In this way, the process of boundary condition setting is divided into the following sub-steps:
\begin{enumerate}
\item Identification of numerical boundary points.
\item Evaluation of the numerical boundary conditions.
\item Embedding in the regular domain $D$.
\end{enumerate}
The details of the above sub-steps are in order. In what follows, we will consider one immersed body. For the multi-object problems, the method can be applied exactly in the same way. Moreover, according to Fig. \ref{fig3}, we assume that the physical velocity boundary conditions ${\bf u}_{\rm pb}$ are given on the physical boundary $\Gamma_{\rm P}$, and the solution is sought in $D\setminus (\Omega\cup \bar{\cal B})$, and the regular domain $D$ is overlaid by a uniform Cartesian grid $(x_i,y_j)$.
\subsubsection{Identification of numerical boundary points}
\begin{figure}[t]
\setlength{\unitlength}{1mm}
\centerline{\includegraphics[width=9cm]{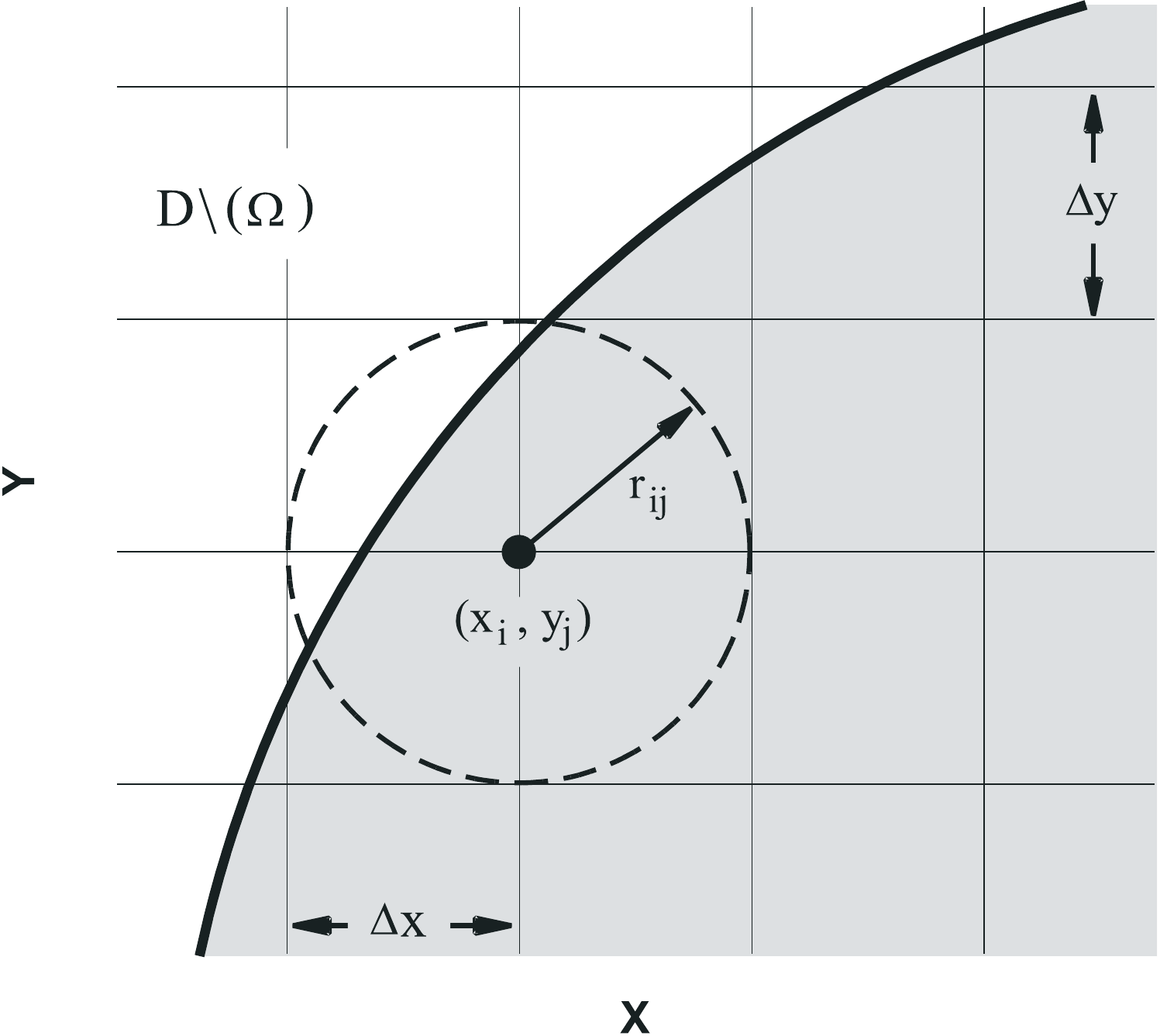}}
\caption{A numerical boundary point defined on a uniform grid.} \label{fig4new3}
\vspace{5mm}
\end{figure}

In the present method, all calculations are performed on a fixed Cartesian grid. Therefore, in the following of our previous work \cite{Sabetghadam}, we define the numerical boundary points, which play the role of Eulerian points in some fluid--solid interaction methods, which use both the Eulerian and Lagrangian points (see {\it e.g.} \cite{Balaras}).\\ \\
{\bf Definition 1.} A numerical boundary point corresponding to the given physical boundary $\Gamma_{\rm P}$, is a point $(x_i,y_j)$ in the Cartesian grid, if and only if
\begin{itemize}
\item[{\bf i)}] $(x_i,y_j)\in (\bar{\Omega}=\Omega\cup\Gamma_{\rm P})$
\item[{\bf ii)}] ${\cal C}_{ij}$ contains at least one point from $D\setminus \bar{\Omega}$
\end{itemize}
where ${\cal C}_{ij}$ is a circle of radius $r_{ij}={\rm min}({\Delta x,\Delta y})$, centered in $(x_i,y_j)$. The definition is illustrated in Fig. \ref{fig4new3} (for a uniform grid), and for some more details one can see \cite{Sabetghadam}.
The set of all numerical boundary points will be called the numerical boundary $\Gamma_{\rm N}$, and that part of the Cartesian grid which is surrounded by $\Gamma_{\rm N}$ will be called the numerical immersed domain $\Omega_{\rm N}$.\\
For the fixed boundary problems, it is just needed to determine the numerical boundaries once for all computations, while for the moving boundary problems they should be updated in the beginning of each timestep with the computational cost of ${\cal O}({\cal N}_{{\bar \Omega}})$, where ${\cal N}_{{\bar \Omega}}$ is the number of grid points in a box, contains the immersed domain ${\bar \Omega}=\Omega_{N}\cup\Gamma_{N}$.
\subsubsection{Approximating the boundary conditions} \label{BCs}
Given the numerical boundary points $\Gamma_{\rm N}$, the next step is to evaluating velocities on these points (will be called ${\bf u}_{\Gamma_{\rm N}}$), such that the physical boundary conditions ${\bf u}_{\rm pb}$ be satisfied approximately. Among several possibilities, an extrapolation method is borrowed from the finite difference--based immersed boundary methods \cite{Sjogreen}, because of its flexibility, and its consistency with our next steps. \\
\begin{figure}[t]
\setlength{\unitlength}{1mm}
\centerline{\includegraphics[width=8cm]{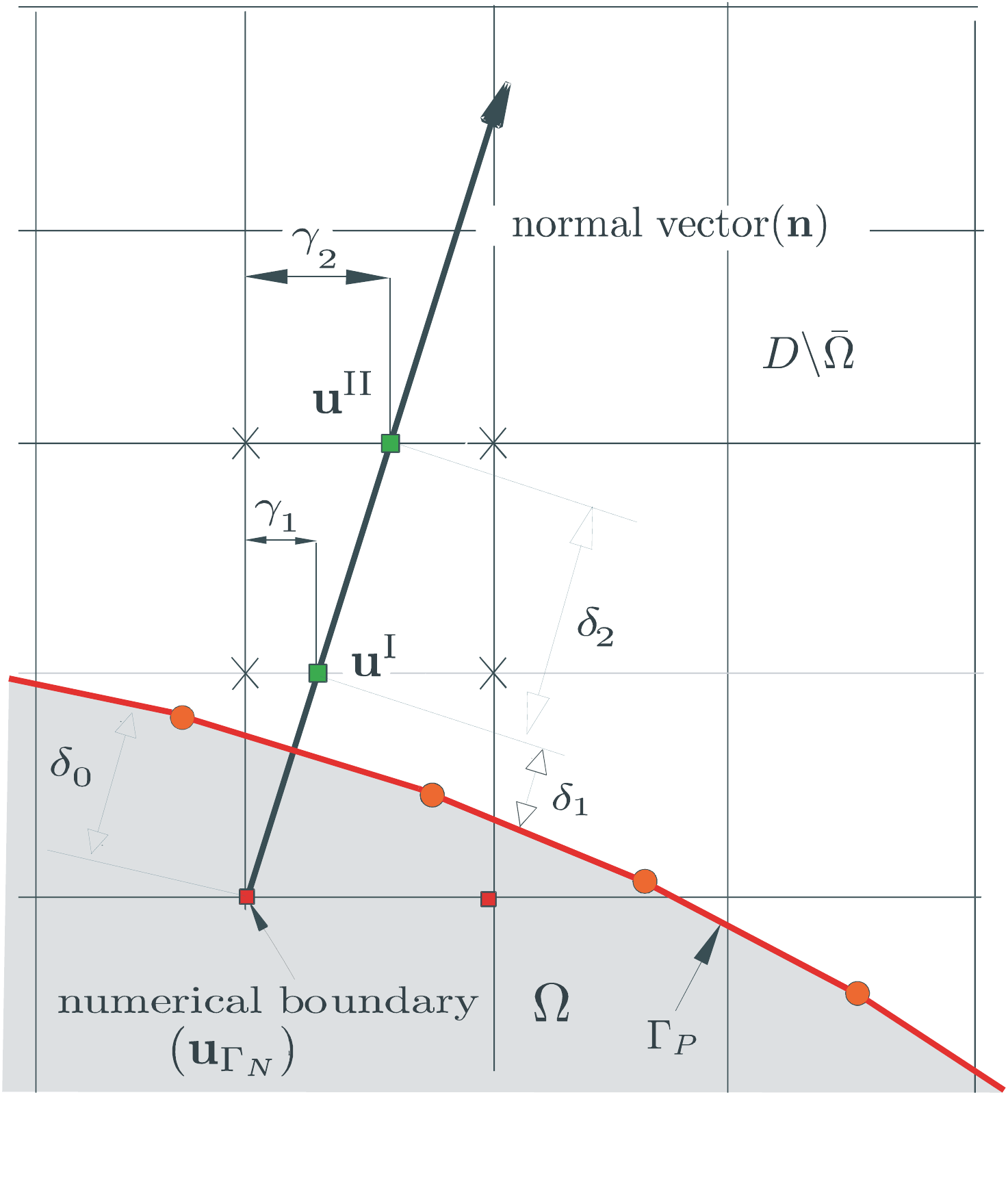}}
\caption{The geometric parameters, and the interpolation points for a second--order boundary condition setting ({\it i.e.}, ${\cal N}_{\rm p}=2$). The auxiliary quantities ${\bf u}^{\rm I}$ and ${\bf u}^{\rm II}$ are obtained from linear interpolations between the neighboring points, which are shown by $\times$.} \label{fig4new}
\vspace{8mm}
\end{figure}
Basically, the method is an order ${\cal N}_p$ polynomial extrapolation, along the local normal direction $\bf n$.  According to Fig. \ref{fig4new}, given the velocity vector ${\bf u}=(u_1,u_2)$, the auxiliary velocities ${\bf u}^{\rm I}$, ${\bf u}^{\rm II}$\ldots (number of them depends on ${\cal N}_p$), are determined from some linear interpolations between the neighboring points on the Cartesian grid.  Then the desired approximation is obtained from
\begin{equation}
{\bf u}_{\Gamma_{\rm N}}={\bf f}_{{\cal N}_p}({\bf u}^{\rm I},{\bf u}^{\rm II},\ldots; {\bf u}_{\rm pb}),\label{extrapolation}
\end{equation}
where ${\bf f}_{{\cal N}_p}$ is the polynomial extrapolation function of order ${\cal N}_p$ , and ${\bf u}_{\rm pb}$ is the given physical velocity boundary condition. The neighboring points which are involved in calculation of the auxiliary velocities are depended on the slope of the normal direction ${\bf n}$ (see \cite{Sjogreen}). In the present paper ${\cal N}_p=0,1,2$ are used in our different test cases, and  also in different regions of each test case, depending on the desired local accuracies.\\
As an example, for a second order extrapolation (that is, ${\cal N}_{\rm p}=2$), according to Fig. \ref{fig4new}, the numerical velocity boundary conditions can be obtained from
\begin{equation}
{\bf u}_{\Gamma_{\rm N}}=\frac{\delta_2(\delta_2+\delta_3)}{(\delta_1+\delta_2)P}{\bf u}_{\rm pb}+\frac{\delta_1(\delta_2+\delta_3)}{\delta_3(\delta_1+\delta_2)}{\bf u}^{\rm I}- \frac{\delta_1\delta_2}{\delta_3 S}{\bf u}^{\rm II},
\end{equation}
where $S=\delta_1+\delta_2+\delta_3$; and ${\bf u}^{\rm I}$, and ${\bf u}^{\rm II}$ are found from linear interpolations.\\
With regard to these extrapolations, the following practical points are worth mentioning:
\begin{itemize}
\item[(1)] Performing these extrapolations showed to be not so easy for complex geometries and moving boundaries. Moreover, for ${\cal N}_p>0$ it may increase the sharpness of the velocity fields, which particularly for the spectral solutions may increase the Gibbs oscillations by triggering the higher wavenumber modes.
\item[(2)] For ${\cal N}_p=0$, it is not any extrapolation indeed, and the boundary condition setting reduces to using a mask function, which substantially simplifies the process; and in many cases, reduces the Gibbs oscillations. However, for the boundaries which are not coinciding with the Cartesian grid, it means zero-order implementation of the boundary conditions, which reduces the global accuracy of the solution, as it will be seen in $\S$ \ref{num_exp}.
\end{itemize}
By substitution of ${\bf u}_{\Gamma_{\rm N}}$ into $\bf u$, an auxiliary modified velocity vector ${\bf u}_{\rm nb}$ obtains, which is identical to $\bf u$ in $D\setminus\Omega_{\rm N}$, and satisfies (approximately) the physical boundary conditions on $\Gamma_P$.\\
\subsubsection{Embedding in the regular domain} \label{extension}
To take advantages of the pseudo-spectral solutions, the auxiliary modified velocity vector ${\bf u}_{\rm nb}$ should be embedded in the regular domain $\bar D$; and in order to minimizing the Gibbs oscillations, the final modified embedded velocities ${\bf u}^{\rm BC}$ should be as smooth as possible. Therefore, referring to Fig. \ref{fig3}, the final modified velocity ${\bf u}^{\rm BC}$ should satisfy the following conditions
\begin{equation}
{\bf u}^{\rm BC}=
\cases{
{\bf u}_{\rm nb} & {\rm on} ${{\bar D}\setminus({\bar{\cal B}}\cup \Omega_{\rm N}})$, \cr
{\bf u}_{\rm pb} & {\rm in} ${\Omega_{\rm N}}$, \cr
{0} & {\rm in} ${\cal B}$. \cr}
\end{equation}
Moreover, smooth transitions from ${\bf u}_{\rm nb}$ to ${\bf u}_{\rm pb}$ (in $\Omega_{\rm N}$), and from $\bf u_{\rm nb}$ to zero (in $\cal B$) are desired, in order to minimizing the Gibbs oscillations. In this context, the following two-step procedure is followed for the sake of flexibility:
\begin{enumerate}
\item {\bf Extension:} Given the extrapolation functions ${\bf f}_{{\cal N}_p}$ (for different points of $\Gamma_{\rm N}$ and $\Gamma_{\cal B}$), the desired extensions of ${\bf u}_{\rm nb}$ (in $\Omega_N$ and $\cal B$), are easily obtainable. Note that these extensions have ${\bf C}^{{\cal N}_p}$ smoothness (in the directions normal to the immersed boundaries), because they are polynomials of order ${\cal N}_{p}$. The extended velocity will be called ${\bf u}^{D}$ (since it is defined on $\bar D$), and note that ${\bf u}^D$ is not necessarily periodic.
\item {\bf Smoothing and periodization:} Finally ${\bf u}^{\rm BC}$ is obtained from smoothing and periodization of ${\bf u}^D$. Among some methods that can be seen in the literature (see {\it e.g.} \cite{B2002, Bueno, Bruno}), we used multiplication of ${\bf u}^D$ by an appropriate window function \cite{Sabetghadam}
\begin{equation}
{\bf u}^{\rm BC}({\bf x})= \varrho({\bf x}) \cdot {\bf u}^{D}({\bf x}),
\end{equation}
where the window function $\varrho({\bf x})$ is constructed such that it is sufficiently smooth in $\Omega_{\rm N}$ and is zero in $\cal B$. To construct the desired window function, we used shifted and scaled error functions, exactly the same as our previous work \cite{Sabetghadam}.
\end{enumerate}
\begin{figure}[t]
\setlength{\unitlength}{1mm}
\centerline{\includegraphics[width=9.5cm]{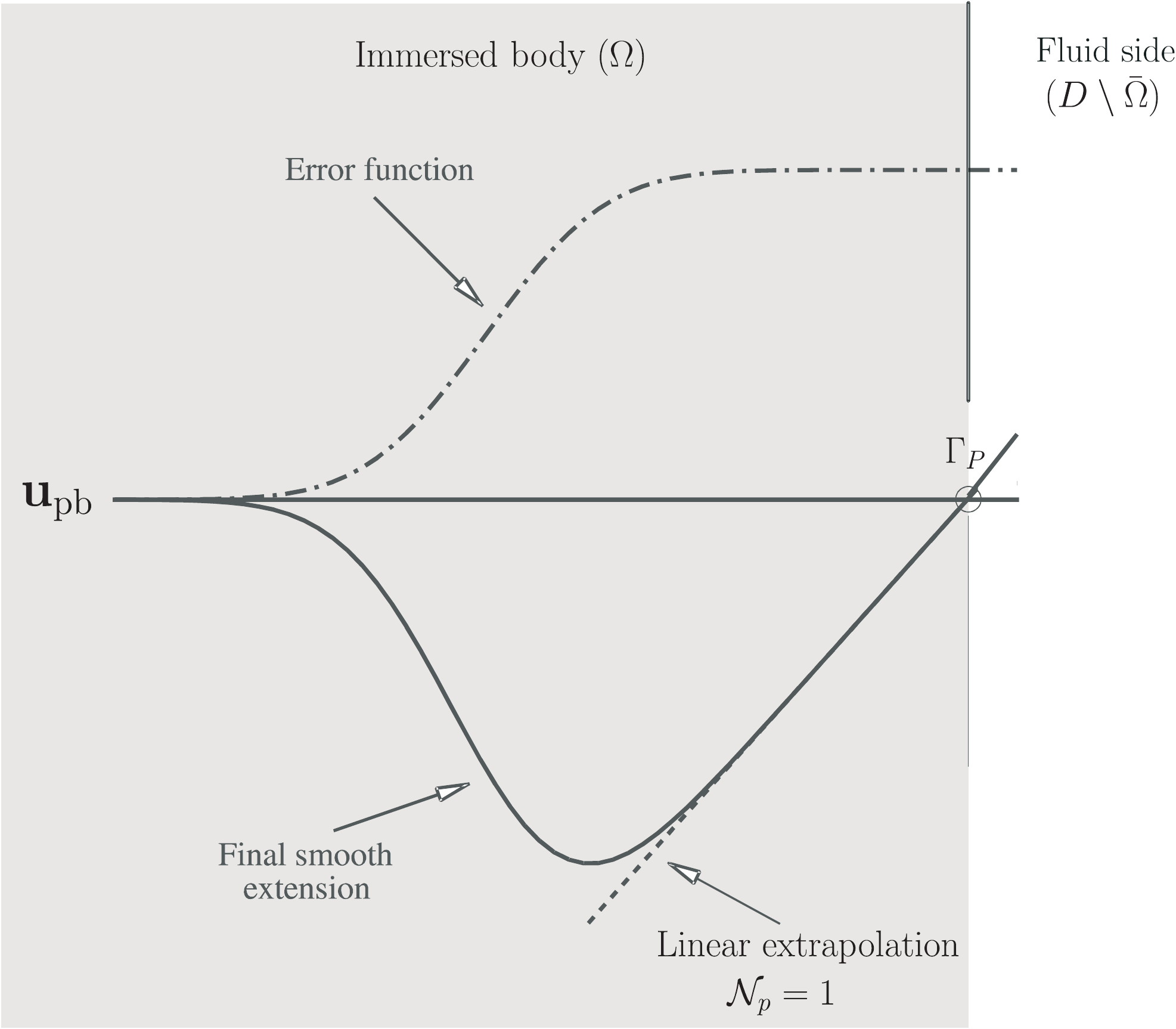}}
\caption{A sectional view of an extended and periodized field for ${\cal N}_{p}=1$. The $u_i^{\rm BC}$ are obtained from multiplication of a linear extension (with a slope which is computed from the internal points adjacent to the physical boundary $\Gamma_p$) by a shifted error function.}
 \label{fig5}
\vspace{5mm}
\end{figure}
A sectional view of an extended and periodized velocity field for linear extrapolation (${\cal N}_{p}=1$) is shown in Fig. \ref{fig5}. The physical boundary condition ${\bf u}_{\rm pb}$ was given on the physical boundary $\Gamma_P$ ; and note that because of the slope of the solution in the vicinity of $\Gamma_{P}$ (since ${\cal N}_P>0$), a margin with ${\bf u} \neq {\bf u}_{\rm pb}$ is appeared in the immersed body, which will be ignored in the solution procedure, like other immersed boundary methods.\\
Apparently, the computational cost of boundary condition setting is of ${\cal O}({\cal N}_{\Omega '})$, where
${\cal N}_{\Omega '}$ is the number of grid points in $\Omega '={\Omega}_{\rm N}\cup{\cal B}$. However, It should be noted that for the boundaries which are coinciding with the Cartesian grid ({\it e.g.} the rectangular boundaries for the Cartesian grids), the extrapolation step can be bypassed.
\subsection{Filtering of the oscillations} \label{Helm}
Although the modified velocities ${\bf u}^{\rm BC}$ are $C^{{\cal N}_p}$ in the directions normal to the immersed boundaries, and they are periodized; however, for the finite Reynolds numbers and complex geometries there is no any guarantees for their global smoothness \cite{Sabetghadam}. Moreover,  the conditioned vorticity $\omega^{{\rm BC}}=\nabla \times {\bf u}^{\rm BC}$ is generally sharper than ${\bf u}^{\rm BC}$. Therefore, as our numerical experiments have confirmed as well, the solution suffers from the Gibbs oscillations almost in all practical situations.  Fortunately, there are some evidence which show that despite these oscillations the solutions are still accurate in the weak sense, and higher orders of pointwise accuracies can be recovered (if desired), using appropriate numerical filters \cite{Gotlib, Keetels, Kolomenskiy}.\\
In the present paper, the Helmholtz filter is used (among many kinds of the numerical filters), because of its simplicity, and since it is aimed to use it just as a {\it postprocessor} in presentation of the results, not for achieving the pointwise accuracies or improvement of the rates of convergence. However, our other numerical experiments (are not presented in this paper, because of lack of any theoretical support), have shown that use of this filter in the solution procedure may reduces the Gibbs oscillations.\\
Given the non-filtered quantity $\phi({\bf x})$, the Helmholtz-filtered quantity $\bar{\phi}({\bf x})$ is obtained from the following convolution
\begin{equation}
\bar{\phi}({\bf x})=(1-\alpha^2 \nabla^2)^{-1}\phi({\bf x}),
\label{helmholtz1}
\end{equation}
where $\alpha$ is the free parameter of the filter, standing for an appropriate length scale (see \cite{Germano}, or the series of works on the $\alpha$ modeling of turbulent flow {\it e.g.} \cite{Foias}). Since the Laplacian operator is diagonal in the Fourier space, the above convolution takes a simple form in the wavenumber space
\begin{equation}
\hat{\bar{\phi}}({\bf k})=\frac{\hat{\phi}({\bf k})}{(1+\alpha^2|{\bf k}|^2)},
\label{helmholtz2}
\end{equation}
where ${\bf k}$ is the wavenumber vector. Since the dimension of $\alpha$ is length, it is convenient to define a more general non-dimensional filtering factor ${\bf C}_{\alpha}$ such that
\begin{equation}
\alpha=\frac{2\pi}{L}\cdot \frac{1}{{\bf C}_{\alpha}{\bf k}_{\rm max}} ,
\label{calpha}
\end{equation}
where $L$ is the length of the solution domain, and
\begin{equation}
{\bf k}_{\rm max}= ||{\bf k}_{\rm max}||_2 =\sqrt{k_{\rm max 1}^2+k_{\rm max 2}^2}~~,
\end{equation}
is the maximum wavenumber, involved in the solution.\\
The above filter is used in the next section, wherever it is needed to remove the Gibbs oscillations, for better presentation of the results.
\begin{figure}[t]
\setlength{\unitlength}{1mm}
\centerline{\includegraphics[width=8cm]{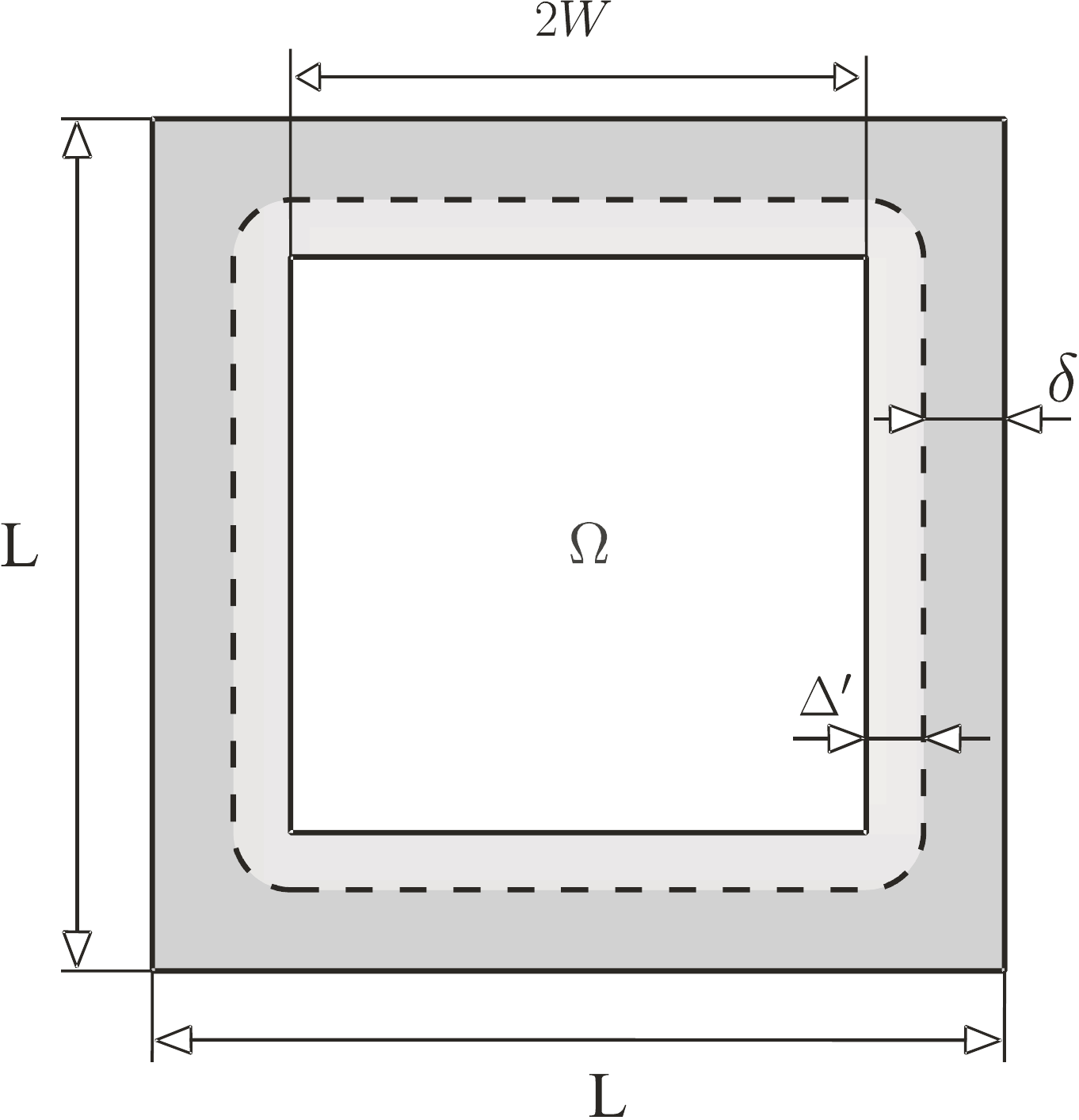}}
\caption{Geometric parameters of the dipole--solid wall collision problem. In order to maximizing the active grid points, $\delta$ is chosen as small as possible, and different smoothness of windowing is obtained by choosing different $\Delta'$ (see also Fig. \ref{fig7}).} \label{fig6} \vspace{5mm}
\end{figure}
\section{Numerical experiments} \label{num_exp}
In order to assess the capabilities of the method, three test cases are analyzed in this section, including the fixed and moving boundaries, as well as given non-zero immersed velocity boundary conditions.
\subsection{Dipole--solid wall collision} \label{dipole-wall}
As a coordinate-coinciding immersed boundary problem, the recently re-analyzed dipole--solid wall collision problem  \cite{Orlandi,Clercx2006} is chosen as our first numerical experiment. In a series of papers, the problem has been analyzed numerically for the normal and oblique collisions and a fairly wide range of Reynolds numbers; via finite difference, Fourier and Chebyshev spectral methods, and the volume penalization of the NSE \cite{Clercx97,Clercx2001,Clercx2002,Clercx2006,Keetels}. In order to make quantitative comparisons ${\rm {\bf Re}}=\frac{UW}{\nu}=1000$ is chosen in this paper, similar to the Keetles {\it et al.} test case \cite{Keetels}.
\begin{figure}[t]
\setlength{\unitlength}{1mm}
\centerline{\includegraphics[width=9cm]{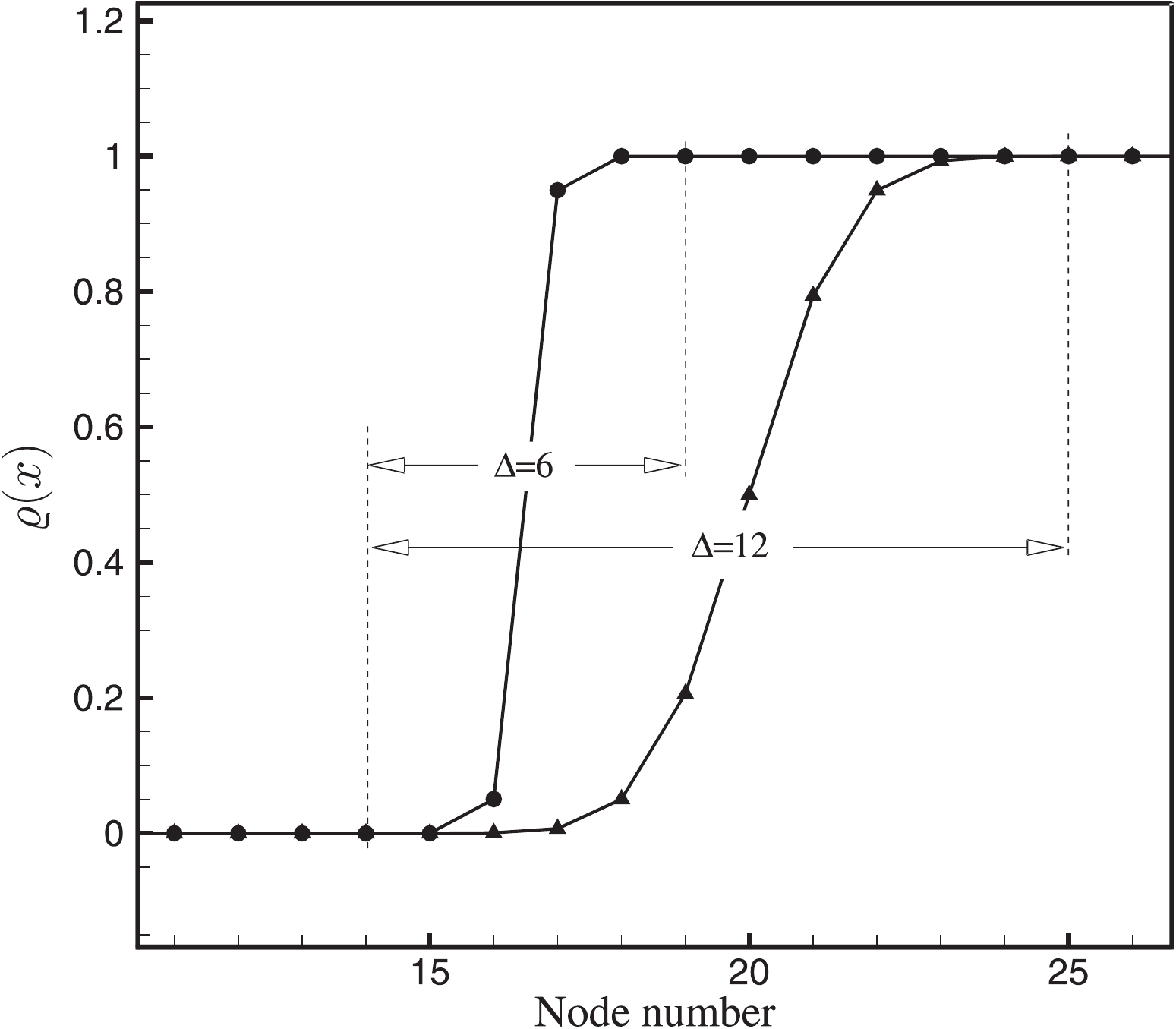}}
\caption{A sectional view of two window functions: a fairly sharp window $\varrho_6({\bf x})$  that $\Delta=\frac{\Delta '}{\Delta x}=6$, and a moderately smooth one $\varrho_{12}({\bf x})$ that $\Delta=12$. The rising distance $\Delta'$ is the distance between the positions that $\varrho=0$ and $\varrho=1$ with the tolerance of ${\cal O}(10^{-14})$.} \label{fig7} \vspace{5mm}
\end{figure}
\subsubsection{The problem setup}
The geometric parameters are illustrated in Fig. \ref{fig6}, where it is chosen $2W=2$ for simplicity (c.f. \cite{Keetels}). The margin $\delta$ should be chosen as small as possible, in order to maximizing the number of active grid points. On the other hand, it should be sufficiently wide in order to satisfy the requirements that are discussed in the appendix. Here $\delta=10\Delta x$ is chosen (after a trial and error process) in all runs. Although fairly acceptable solutions were obtained for smaller $\delta$, solutions exhibited more Gibbs oscillations.\\
\begin{figure}[t]
\setlength{\unitlength}{1mm}
\centerline{\includegraphics[width=10cm]{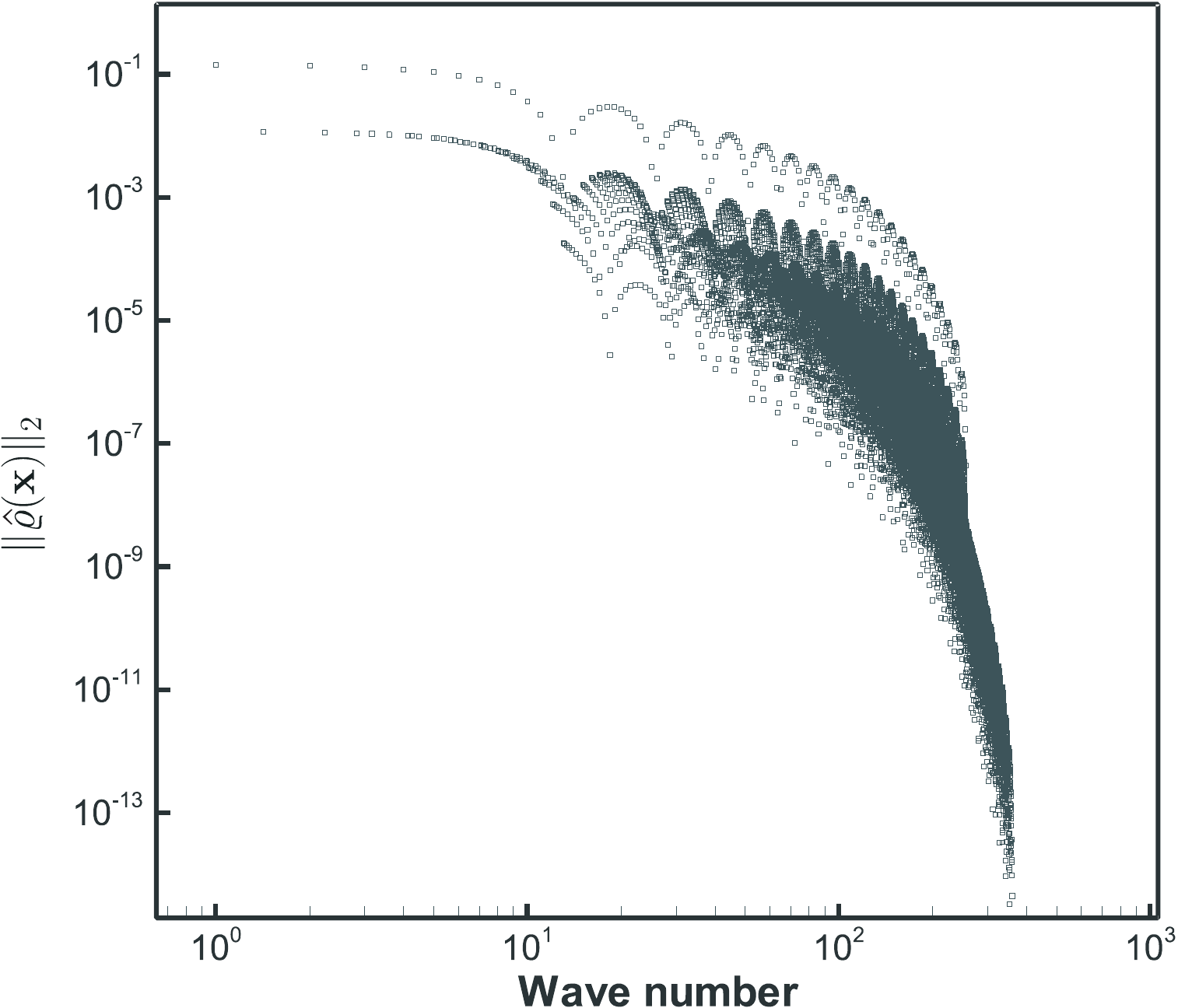}}
\caption{Scatter diagram of the window function $\varrho_{12}({\bf x})$. The exponential rate of decaying, especially in the higher wavenumber, is noticeable (c.f. \cite{Boyd05}). } \label{fig7new} \vspace{5mm}
\end{figure}
As it was mentioned in \S \ref{extension}, for the coordinate-coinciding boundaries, the extrapolation step can be bypassed, and therefore, the boundary condition setting is reduced to merely extension and windowing. In the present test case ${\cal N}_p=1$ is chosen; and various numerical experiments were performed on a fairly wide range of smoothness of the window functions. The results of two typical ones, that is, a sharp window with $\Delta=\frac{\Delta'}{\Delta x}=6$ (henceforth will be called $\varrho_6({\bf x})$), and a moderately smooth one $\Delta=12$ (which will be called $\varrho_{12}(\bf x)$), are reported here. A sectional view of these window functions are shown in Fig. \ref{fig7}. It is worth mentioning that although a square, with sharp corners, is not an ideal shape for a window function; sufficiently smooth square-shaped window functions are obtainable. To emphasize this issue, the scatter diagram of the absolute values of the Fourier coefficients for $\varrho_{12}({\bf x})$ is shown in Fig. \ref{fig7new}. As one can see, the exponential rate of decaying is observable in the higher wavenumber, which can be interpreted as evidence of smoothness of the window function \cite{Boyd05}.\\
In the following of \cite{Keetels}, the initial condition was constructed by summation of two identical mono-poles
\begin{equation}
\omega_0=\omega_e[1-(\frac{r}{r_0})^2]\exp[-(\frac{r}{r_0})^2],
\end{equation}
placed at $\{(x_1,y_1),(x_2,y_2)\}=\{(0,0.1),(0,-0.1)\}$, in which $r_0=0.1$, and $r=\sqrt{x^2+y^2}$. Furthermore, to make quantitative comparisons, the total kinetic energy ${\Bbb E}(t)$, and the total enstrophy ${\Bbb Z}(t)$ are defined as
\begin{equation}
{\Bbb E}(t)=\frac{1}{2}\int_{\Omega}{\bf u}^2({\bf x},t)d{\Omega} ; \quad \quad {\Bbb Z}(t)=\frac{1}{2}\int_{\Omega}\omega^2({\bf x},t)d{\Omega}.
\end{equation}
The main physical parameters of the initial condition are given in Tab. \ref{table1}.\\
\begin{table}[t]
\caption{The physical parameters of the initial condition of the dipole--solid wall collision problem.} \vspace{5mm} \label{table1}
\begin{center}
\begin{tabular}{c c c c c c}\hline \hline
$\omega_e$ & $\nu$ &  $U$  & $\mathbb{E}$ &  $\mathbb{Z}$  & $\rm{{\bf Re}}$ \cr \hline
 299.528385375226&~~ 0.001 ~~&~~1.0~~&~~ 2.0~~ &~~800~~ &~~1000~~\cr \hline \hline \cr
\end{tabular}
\end{center}
\end{table}
The solutions were performed on a $512^2$-points uniform grid and the calculations were de-aliased using the $\frac{3}{2}N$--rule. The explicit fourth-order Runge--Kutta method with a constant timestep $\Delta t=10^{-4}$ was used for time integration. By definition of the {\bf CFL} number as
\begin{equation}
{\rm {\bf CFL}}=||{\bf u}||_{\infty}\cdot \frac{\Delta t }{\Delta {\bf x}},
\end{equation}
in which $||{\bf u}||_{\infty}$ is the maximum velocity magnitude at each time instant, our calculations showed that ${\rm {\bf CFL}}<0.5$ for all times, and all runs.
\begin{figure}[t]
\setlength{\unitlength}{1mm}
\centerline{\includegraphics[width=14cm]{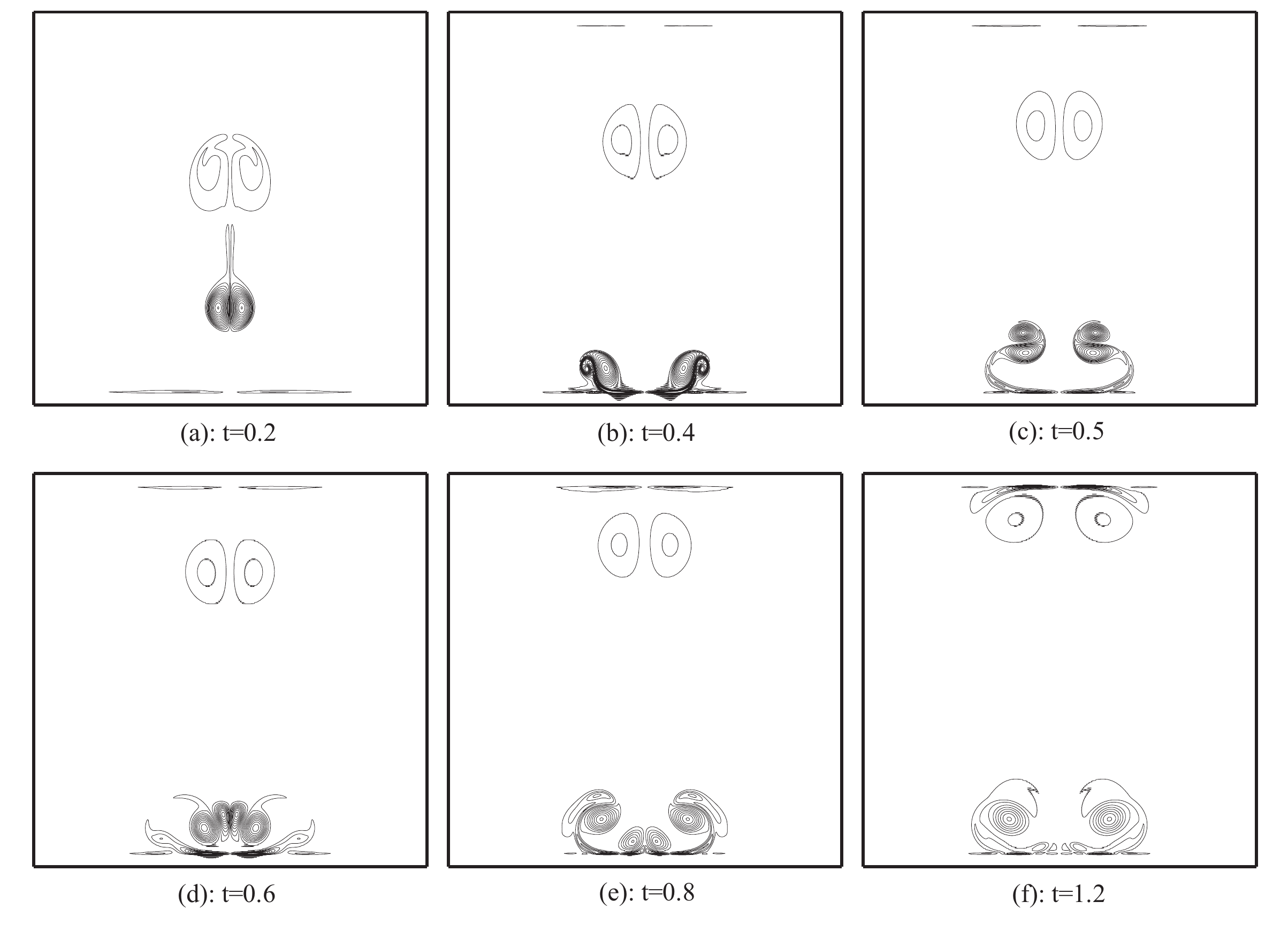}}
\caption{Contour plots of the vorticity field for {\bf Re}=$1000$, obtained from application of the proposed method on a uniform $512^2$-point grid with constant timestep $\Delta t=10^{-4}$ (amounts to {\bf CFL}$<$0.5 for all times). The contour levels (-270,...,-50,-30,-10,10,30,50,...,270), and the time instants are chosen similar to Keetels {\it et al.} \cite{Keetels}.} \label{fig8} \vspace{5mm}
\end{figure}
\subsubsection{The captured physics}
The contour plots of vorticity fields at a number of time instances are presented in Fig. \ref{fig8} for $\varrho_6({\bf x})$ windowing.  The time instances are chosen similar to Keetels {\it et al.} \cite{Keetels} for comparisons. Obviously, the no-slip and the no-penetration conditions, and all phenomena of the dipole--solid wall collision are observable (c.f. \cite{Keetels} or \cite{Clercx2006}).
\begin{figure}[t]
\setlength{\unitlength}{1mm}
\includegraphics[width=0.48\textwidth]{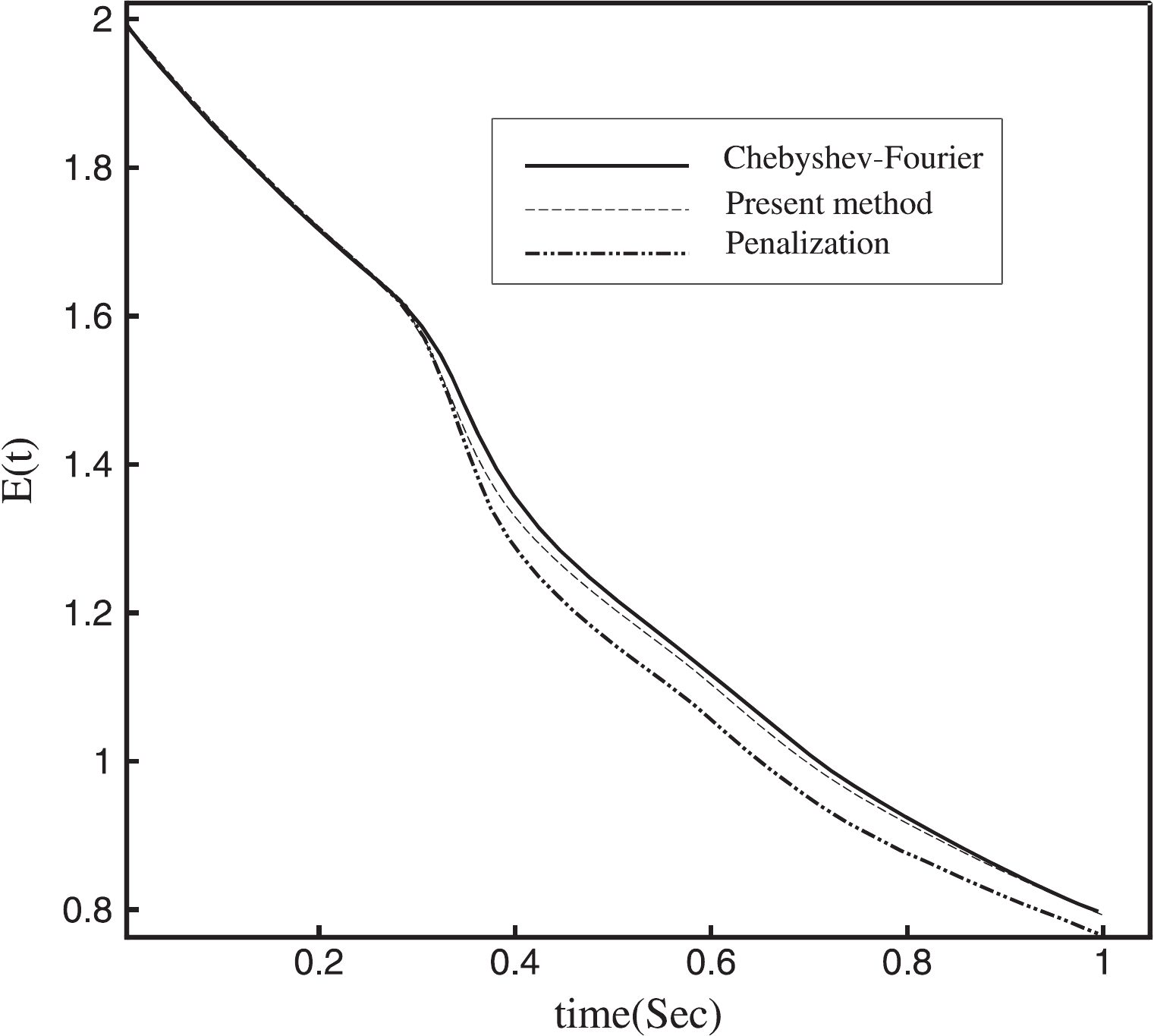}
\includegraphics[width=0.48\textwidth]{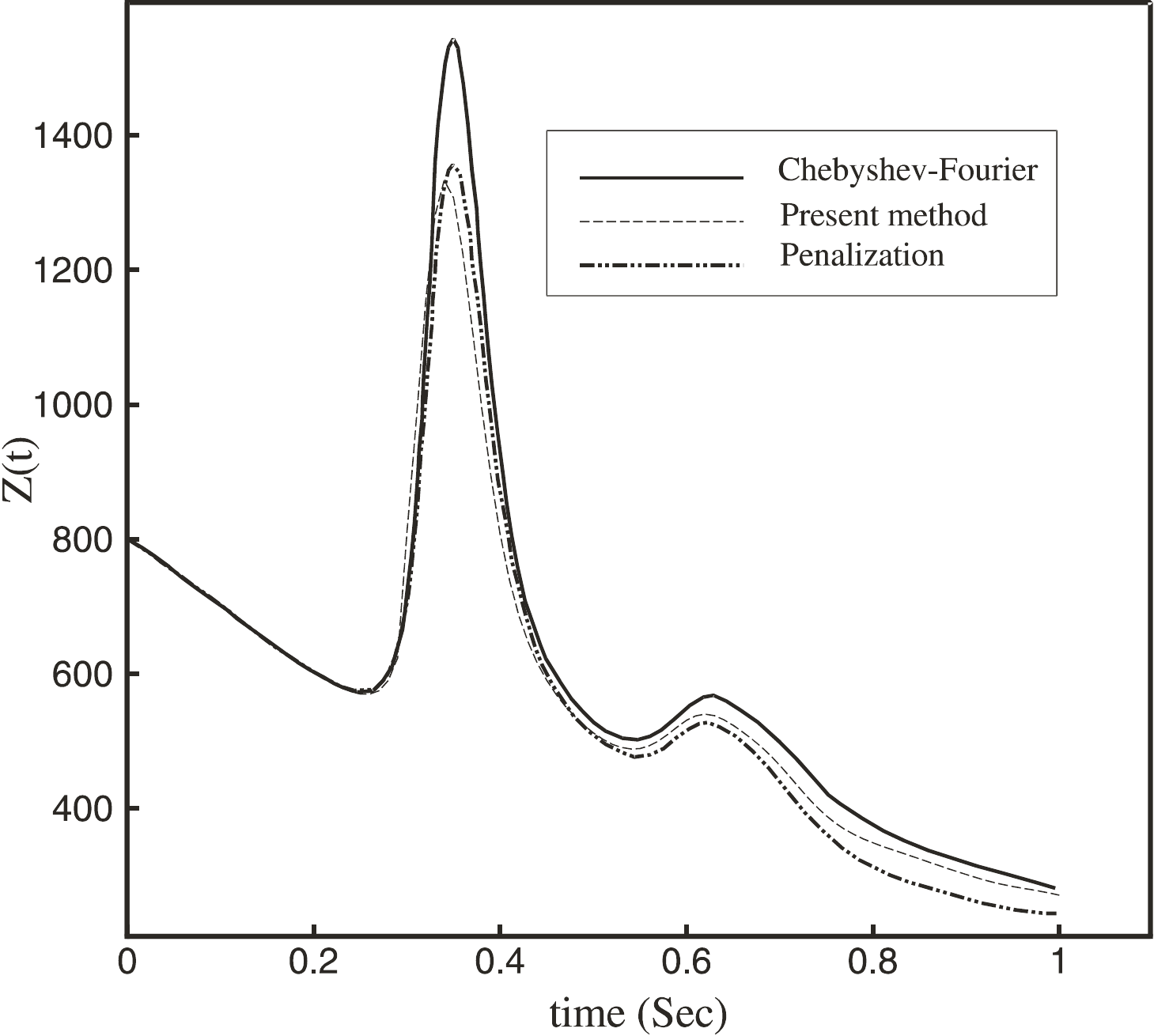}
\caption{Comparison of the total kinetic energy ${\Bbb E}(t)$ and enstrophy ${\Bbb Z}(t)$ of the present method with the Chebyshev-Fourier method and the volume penalized NSE \cite{Keetels}.} \label{fig9} \vspace{5mm}
\end{figure}
More quantitative comparisons can be made by observation of time histories of the total kinetic energy and enstrophy. In Fig. \ref{fig9} these quantities are presented together with the results of Fourier--Chebyshev method and the volume penalization method (these results are extracted directly from \cite{Keetels}). As one can see in the left panel, decay of the total kinetic energy shows a small deviation from the Chebyshev--Fourier results after the first collision. However, this deviation is compensated in the proceeding times, such that after about $t=1$, it is vanished. In the other words, the dissipation rates have been over-predicted during the first collision, and under-predicted for the proceeding times (in comparison to the Fourier--Chebyshev solution). In our opinion, these differences can be interpreted with regard to the smoothness of the solution (which is mainly depended on the grid resolution and sharpness of the windowing, for a fixed Reynolds number), and the order of implementation of the boundary conditions ${\cal N}_p$. In fact, the first extra-dissipation in the collision time interval is supposed to be related to appearance of the Gibbs oscillations, which transfers a part of kinetic energy into the higher wavenumber. It is this high-wavenumber energy part which will be dissipated, faster than it should, in the proceeding times. On the other hand, the under-predicted dissipation rates in the next times is presumed to be a consequence of smooth walls, instead of sharp boundaries. Unfortunately, direct verification of these speculations found to be so difficult in practice, because the smoothness of the solution and the order of boundary conditions are not completely independent (see discussions in $\S$ \ref{BCs} for some more details).\\
\begin{figure}[b]
\setlength{\unitlength}{1mm}
\includegraphics[width=0.48\textwidth]{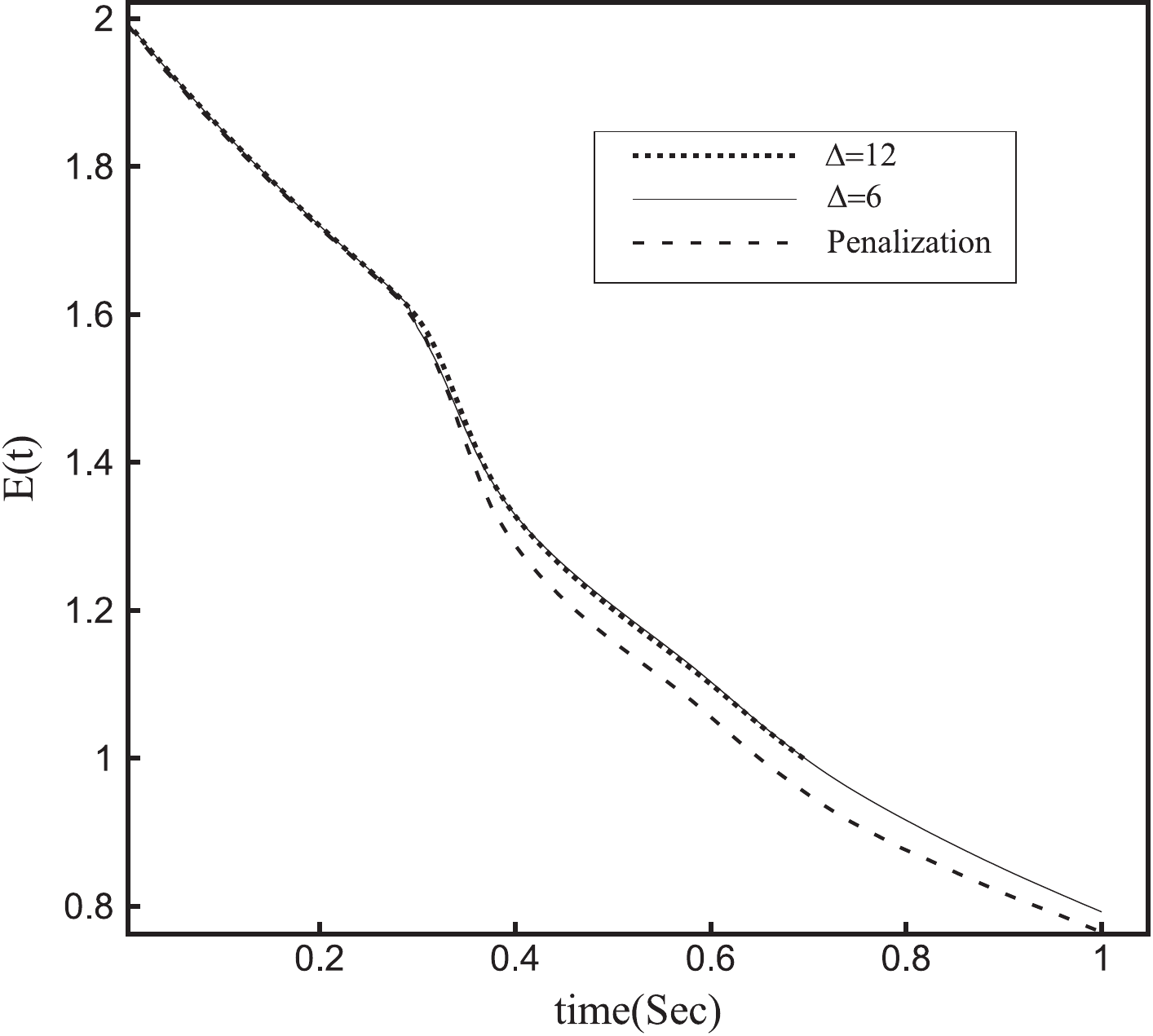}
\includegraphics[width=0.48\textwidth]{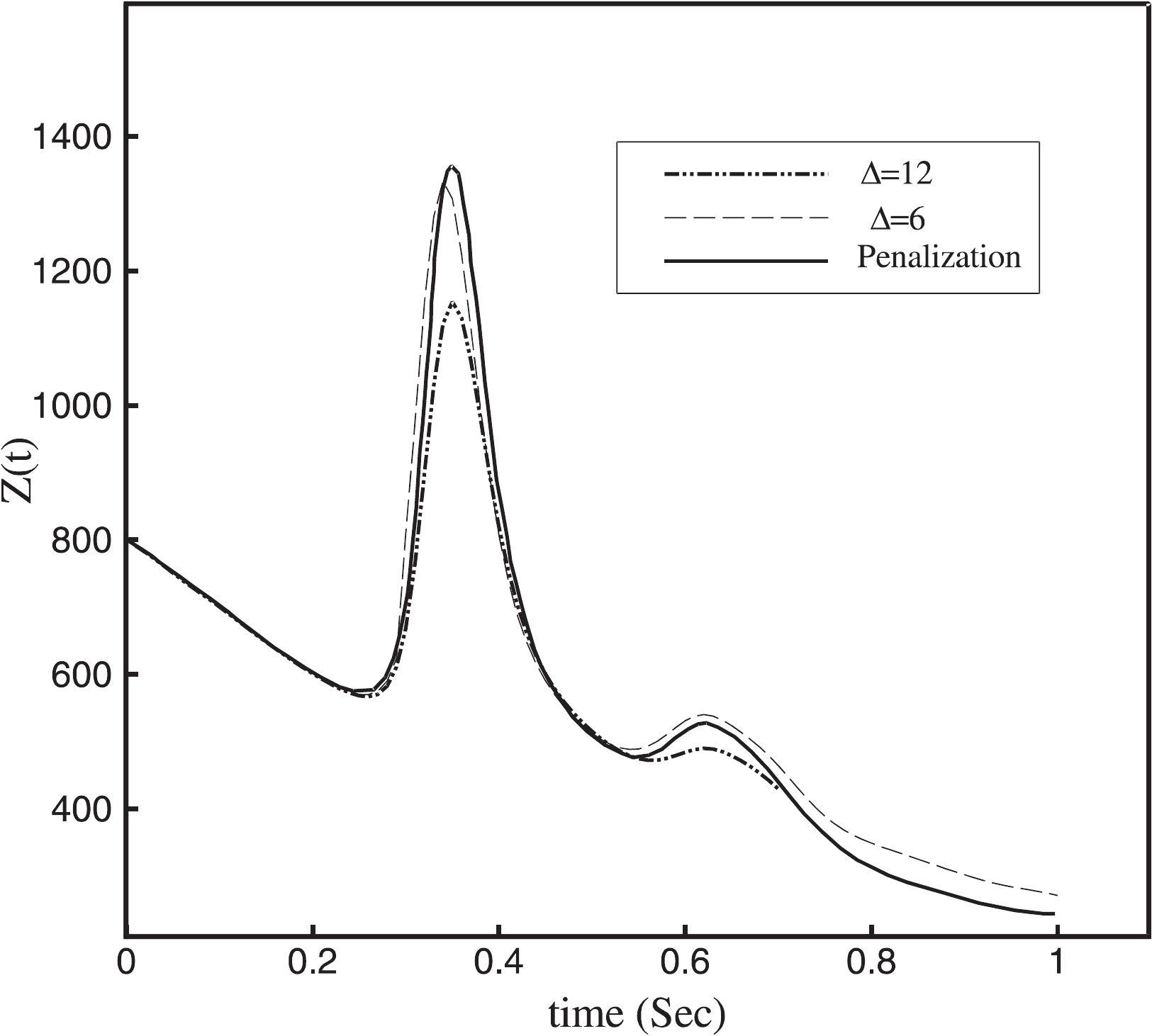}
\caption{Time histories of the kinetic energy and total enstrophy for two window functions (with different sharpness $\varrho_{6}({\bf x})$ and $\varrho_{12}({\bf x})$), are compared with the results of the penalization method. The sharper window function $\varrho_{6}({\bf x})$ resulted in more dissipation and bigger total enstrophy than the smooth window function $\varrho_{12}({\bf x})$. } \label{fig11}
\vspace{5mm}
\end{figure}
In addition to the results of Fourier--Chebyshev method, the results of penalized NSE are presented in Fig. \ref{fig9} as well. Obviously the penalization method showed more dissipations than our method. Again we suppose that it is a consequence of sharpness of the generated boundary layers in the penalized NSE (with ${\cal O}(\sqrt{\epsilon\nu})$ thickness  \cite{Carbou,Keetels}), which is sharper than the window functions in the present method (with maximum sharpness of ${\cal O}(N^{-1})$). Therefore, it is not surprising that the penalization method is more dissipative than the present method in general.\\
On the other hand, in calculation of the total enstrophy (see the right panel of Fig. \ref{fig9}), this sharpness of the penalized NSE, helped the method in generation of the maximum possible vorticity in the Fourier spectral method (that is, the mean value of the vorticity in the vicinity of the discontinuities, as it is argued in \cite{Keetels}); while smoothness of the present method resulted in an under--predicted vorticity generation.\\
For more verification of the effects of smoothness of the windowing process, time histories of the total energy and enstrophy for $\varrho_{6}({\bf x})$ and $\varrho_{12}({\bf x})$ are compared with the penalization method in Fig. \ref{fig11}. Obviously the dissipation rate has changed slightly, while the vorticity generation has affected drastically, especially in the collision times.\\
As a conclusion, our results showed that for this grid-coinciding immersed boundaries, sharper window functions yielded better implementation of the boundary conditions, but more dissipation rates.
\subsubsection{Spatial and temporal rates of convergence} \label{conv_rate}
It is a well-known fact that the rate of convergence of a spectral solution is highly depended on the smoothness of the solution. Particularly, for the non-smooth fields that the Fourier series are suffered from the Gibbs oscillations, the ${\cal L}^{p}$ rate of convergence reduces to about ${\cal O}(1)$. Fortunately, despite this poor ${\cal L}^p$ rate of convergence, it has been shown that high orders of pointwise convergence can be recovered by use of some appropriate numerical filters \cite{Gotlib, Keetels, Kolomenskiy}; although the rates of convergence are depended on the order of the numerical filter this time (confirming that loss of uniform convergence does not mean necessarily loss of the accuracy in the spectral solutions). In this section, to observe the rate of convergence of the method separately, the results are reported without any kind of filtering. However, like some other spectral methods  \cite{Keetels, Kolomenskiy}, these rates can be improved using the numerical filters.\\
\begin{figure}[t]
\setlength{\unitlength}{1mm}
\centerline{\includegraphics[width=8cm]{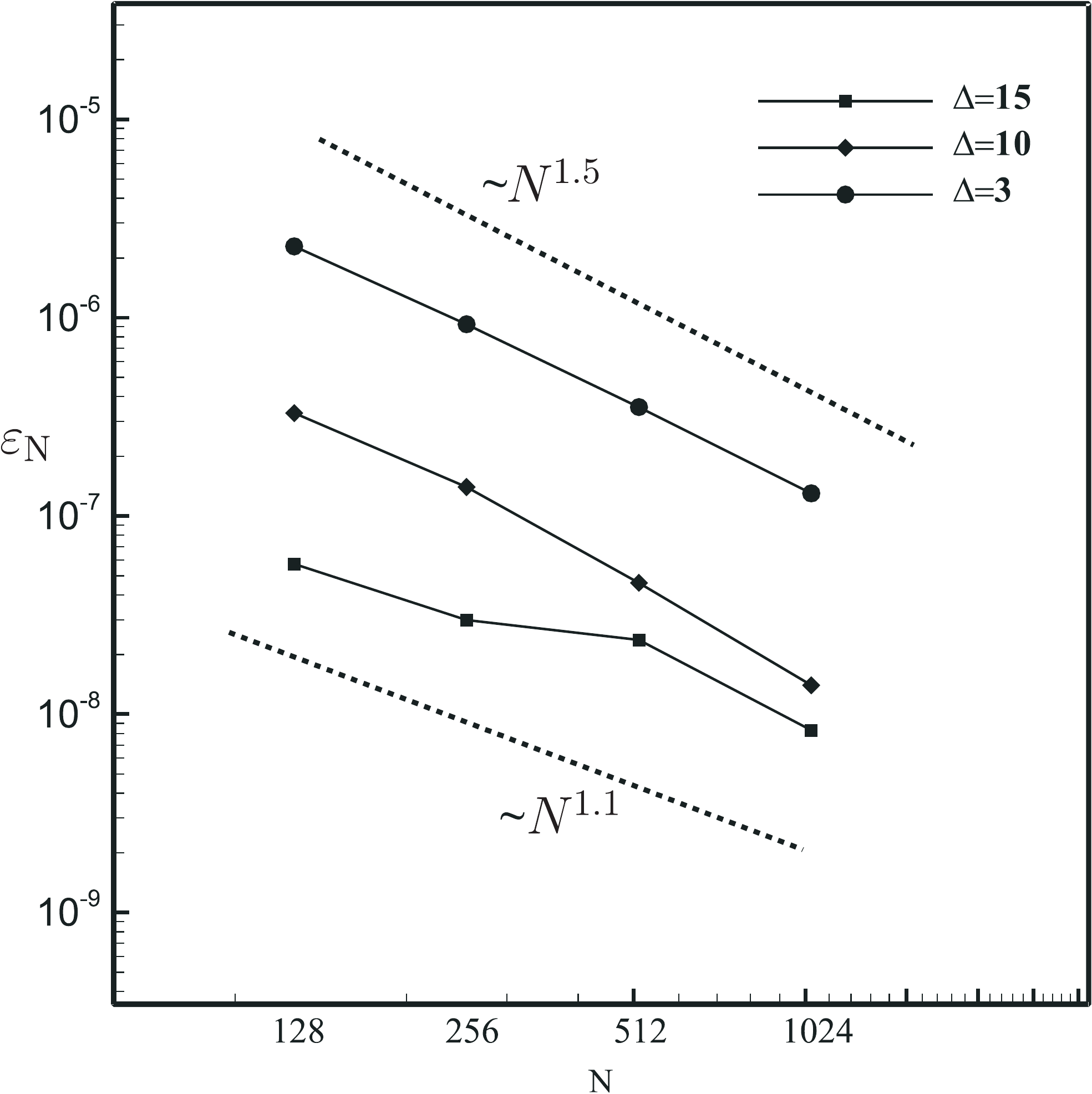}}
\caption{The rates of grid convergence of the vorticity field for three different window functions at $t=0.35$. The rate of convergence approaches to $N^{-1}$ as the sharpness of the window function increase.} \label{fig12}
\vspace{5mm}
\end{figure}
For the initial times that the effects of the solid walls are negligible, the exponential rates for the grid convergence are easily observable for the regions far from the walls. Therefore, merely the  rates of grid convergence in the collision time ($t=0.35$ in particular) in the near-wall region are discussed here. With this regard, the rates of convergence of the vorticity field for three different window functions are presented in Fig. \ref{fig12}.
All solutions were begun from the aforementioned initial conditions, and the solution were continued until $t=0.35$. Similar time step sizes were used for all runs, corresponding to ${\rm  {\bf CFL}}=0.5$ for the finest grid (that is, $2048^2$-point grid). The first-order boundary condition setting ${\cal N}_p=1$ was used, and the normalized ${\cal L}^2$ norms of errors
\begin{equation}
\varepsilon_{\rm N}=\frac{||\omega_{2048}-\omega_{\rm N} ||_{2}}{||\omega_{2048}||_{\infty}},
\end{equation}
were calculated in whole of the flow domain $\Omega$. As one can see, the rates of convergence are approached to ${\cal O}(N^{-1})$ by increasing the sharpness of the window functions.\\
\begin{figure}[t]
\setlength{\unitlength}{1mm}
\includegraphics[width=0.48\textwidth]{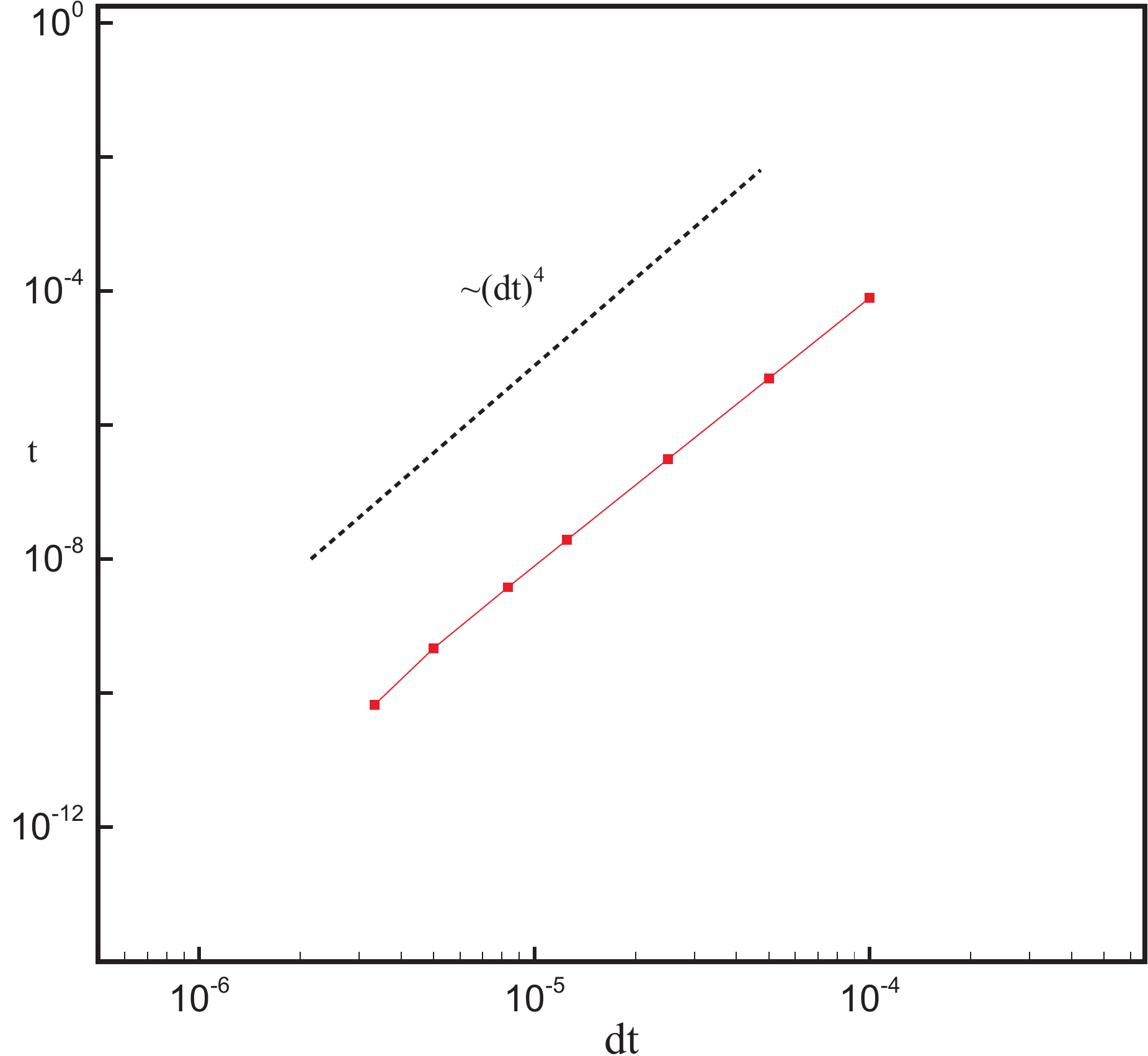}
\includegraphics[width=0.48\textwidth]{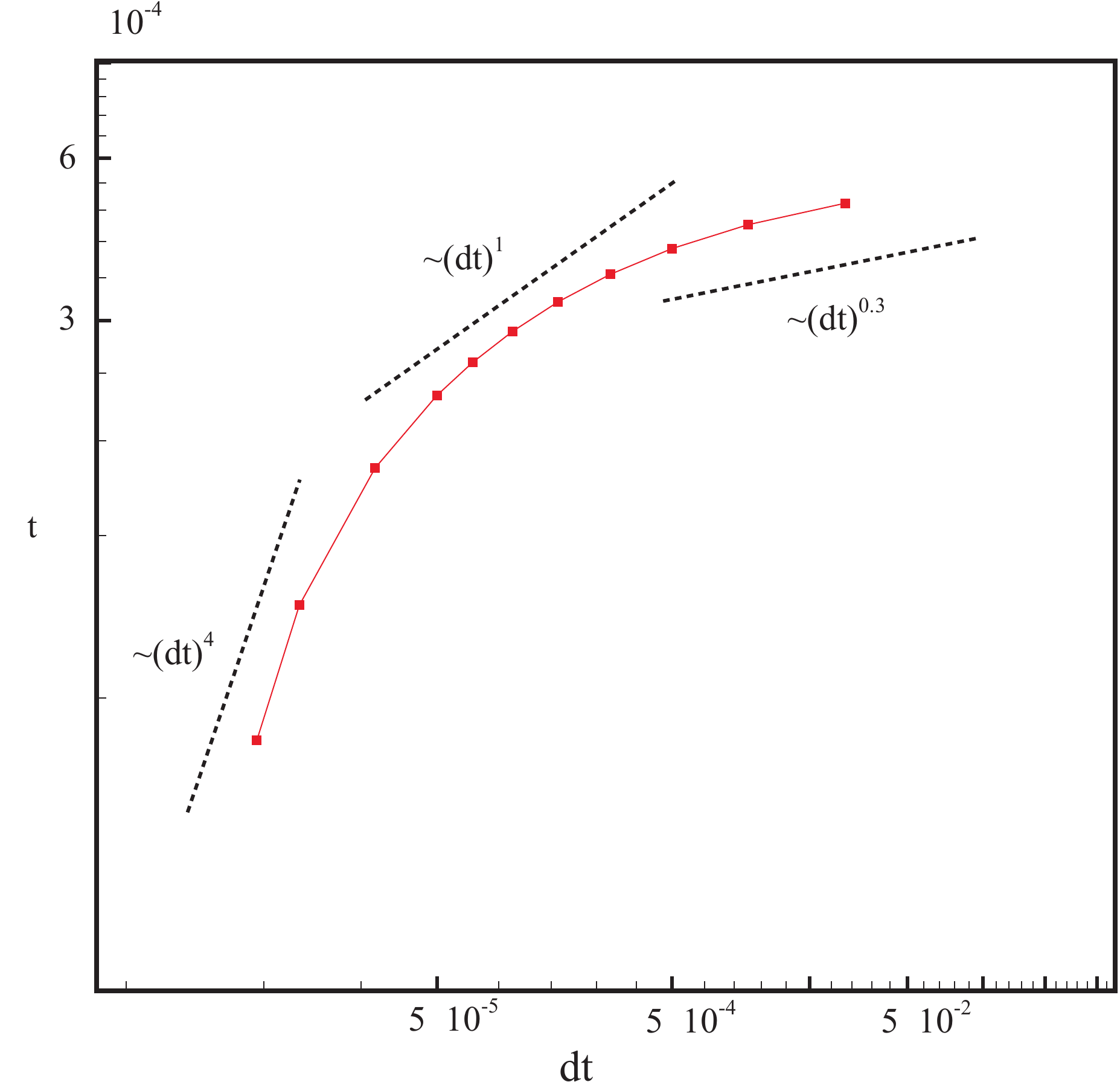}
\caption{The temporal rates of convergence for the no-wall flow (left), and the flow under the influence of the walls (right). The rate of convergence for wall flow is depended on the timestep sizes. However, the maximum rate of convergence of the time integration method ({\it i.e.}, ${\cal O}(dt^4)$) was achieved.} \label{fig13}
\vspace{5mm}
\end{figure}
In calculation of the temporal rates of convergence, and in order to reduce the Gibbs oscillations, a rather smooth window function $\varrho_{12}({\bf x})$ was used on a $512^2$-point grid, and different timestep sizes. At first, we consider the temporal rate of convergence of the no-wall flow. With this regard, we examine the predicted flow at $t=0.1$, for which the effects of the solid walls are approximately negligible. The results are presented in the left part of Fig. \ref{fig13}, which shows the maximum rate of convergence of the fourth-order Runge--Kutta method. Achieving the maximum rate of convergence of the time integration method, primarily shows complete decomposition of the kinematics and dynamics of the flow in the vorticity--velocity formulation of the NSE (as it was mentioned in $\S$ \ref{s2}).\\
Then we repeated the calculations for $t=0.4$, where the flow is under the influence of the solid wall boundary conditions. The results are presented in the right side of Fig. \ref{fig13}. With regard to this figure, the following points are noticeable:
\begin{itemize}
\item[{\bf (i)}]  Our calculations have not shown a constant rate of convergence, rather, the rate of decaying increased by decreasing the timestep sizes. Similar behavior were observed in our other numerical experiments on different test cases and a fairly wide range of timestep sizes. It seems that this behavior of the errors is a consequence of the method of implementation of the boundary conditions, and our insistence on the explicit formulation. More or less similar non-constant temporal rates of convergence have been obtained previously in some other embedded boundary methods \cite{Ditkowski, Abarbanel1, Abarbanel2}. Although absence of a constant rate of convergence is a drawback for a method in general, achievement of the highest rates of convergence of the time integration method (that is, the fourth-order), can be seen as a merit of the method.
\item[{\bf (ii)}]  Although for the sufficiently smooth solution fields the fourth order rate of convergence was obtained, our other numerical experiments have shown that for non-smooth solutions (obtained by {\it e.g.} sharper windowing), the Gibbs oscillations limits again the temporal rates of convergence to ${\cal O}(\Delta t)^{m}$ were $1.1<m<1.3$. Filtering of the solution yields higher rates which are depended mainly on the order of the filter.
\end{itemize}
\begin{table}[b]
\caption{Computational costs for the classical pseudo-spectral (CPS), and the proposed Fourier immersed boundary method (FIBM); for each sub-step of fourth-order Runge--Kutta method. The FFTs are divided into the FFTs on the padded domain (P), and the non-padded domain (N--P).} \label{tab1} \vspace{2mm}
\begin{tabular}{lccccc} \hline \hline
Method   &   N        & ~~~\#FFT (P)&~~~\#FFT (N--P)  &~~~ CPU time &~~~ Memory \\ \hline
FIBM       & $512^2$    &     16         & 4                & 0.947 {Sec.}     & 105 MB  \\
CPS      & $512^2$    &     16         & 0                & 0.905 {Sec.}     & 105  MB \\ \hline \hline
\\
\end{tabular} \label{comp_costs}
\end{table}
\subsubsection{Performance of the method}
The computational costs of the present method in comparison to the classical pseudo-spectral method is given in Tab. \ref{comp_costs}. In this table, the data are presented per boundary condition setting (each Runge--Kutta sub-step). Note that the de-aliasing process (which means $\frac{3}{2}N$ zero-padding), is just needed in obtaining ${\rm {\bf N}}_1$ and ${\rm {\bf N}}_2$ in box {\bf (PS)} in  Fig. \ref{fig4}. Therefore, the FFTs in the boundary condition settings are performed on the original non-padded $N$-point grid, not on the padded grid (with the benifit of increasing the efficiency of the method). To highlight the issue, the padded and non-padded FFTs are separated in Tab. \ref{tab1}. As one can see, implementation of the immersed boundaries needs just four non-padded FFTs, for each sub-step. The CPU times and required memories are obtained for a 3GHz Intel Pentium 4 system. Both methods used the same FFT subroutine and optimization levels, and the CPU times are in average. Apparently, the method is really efficient with less than 5\% additional CPU time and without any sensible increasing in the required memory, in comparison to the Fourier pseudo-spectral method. Such a performance, in addition to robustness of the windowing procedure, makes the method ideal for the moving boundary, multi-object\ldots problems, which some examples can be seen in the sequel.
\begin{figure}[t]
\setlength{\unitlength}{1mm}
\centerline{\includegraphics[width=8cm]{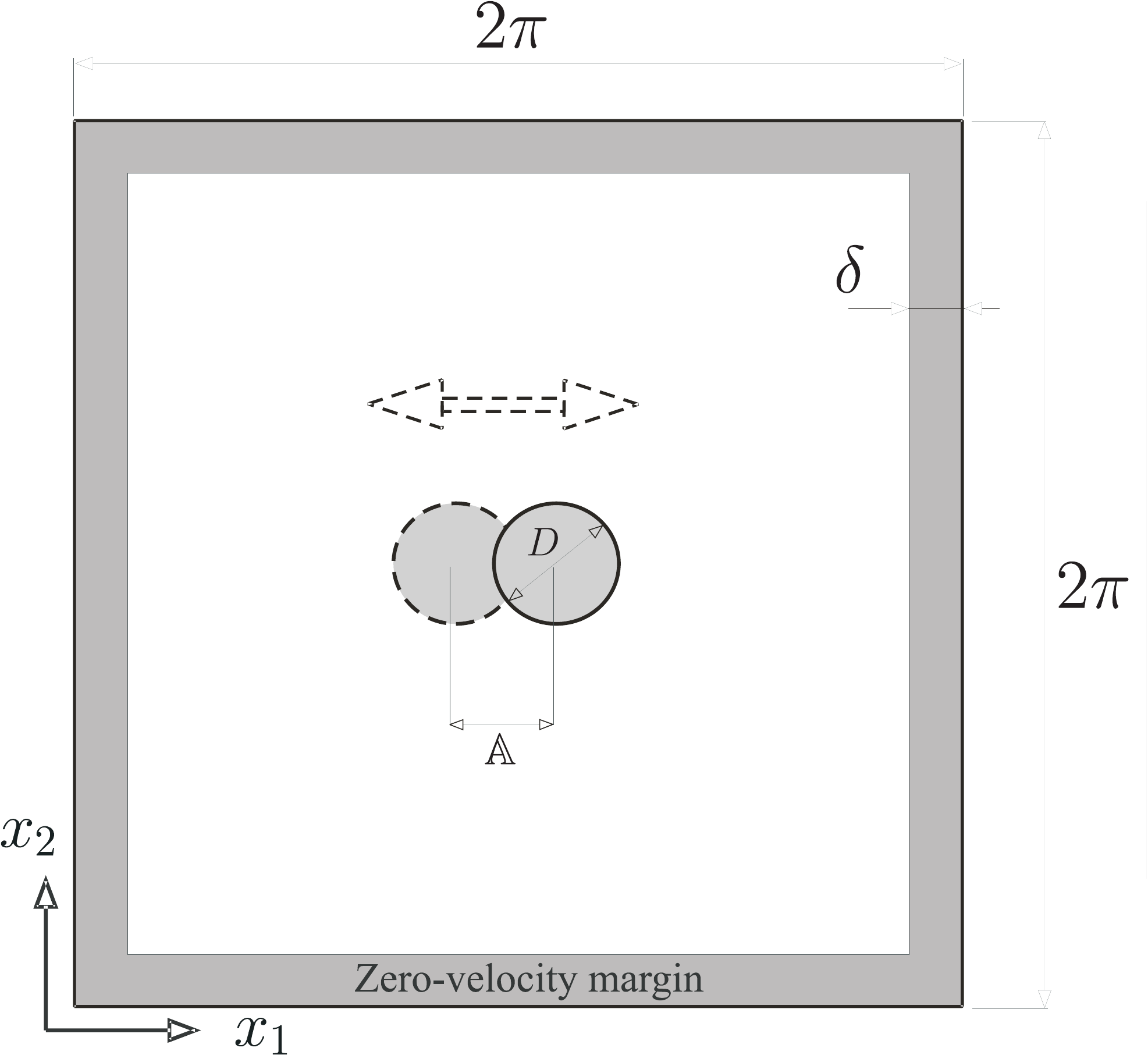}}
\caption{Geometric parameters of the oscillating circular cylinder problem}
\label{fig1cyl}
\vspace{5mm}
\end{figure}
\subsection{Oscillating circular cylinder in fluid at rest}
Undoubtedly, one of the most attractive features of the immersed boundary methods is the possibility of simple and efficient treatment of the complex and moving boundary problems. In the present section, to show the capability of the method in simulation of the moving boundary flow, the problem of in-line oscillating cylinder in a quiescent fluid is examined.
\subsubsection{The problem setup}
The problem has been analyzed several times experimentally \cite{Dutsch}, as well as numerically \cite{choi, Balaras, McDonough}, in a fairly wide range of involved physical parameters; which are the Reynolds number ${\rm{\bf{Re}}}=UD\nu^{-1}$, and the Keulegan--Carpenter number ${\rm{\bf{KC}}}=U(fD)^{-1}$. In these definitions, $U$ is the maximum velocity, $f$ is the oscillations frequency, and $D$ is the cylinder diameter. In the present study, to facilitate assessment of the results, the well-documented combination $({\rm {\bf Re}},{\rm{\bf KC}})=(100,5)$ is considered, which the previous experiments have shown that two-dimensional calculations are rather justifiable \cite{Dutsch}.\\
A simple harmonic oscillation in the $x_1$ direction is exerted to the cylinder center as
\begin{equation}
x_{1c}(t)={\Bbb A}\sin(2\pi f t),
\end{equation}
where $\Bbb A$ is the amplitude of the oscillations. It can easily be shown that for such harmonic oscillations ${\rm{\bf{KC}}}=2 \pi{\Bbb A}D^{-1}$. Fig. \ref{fig1cyl} illustrates the problem setup, and Table (\ref{table_cyl}) summarizes the main physical and numerical parameters of the solution.\\
\begin{table}[t]
\centering
\caption{The physical as well as numerical parameters, used in the solution of the oscillating cylinder problem.}\vspace{5mm}
\begin {tabular}{cccccccc}
\hline \hline
 $D$ & $f$ & ${\Bbb A}$ & $\nu$ & $\delta/{\Delta x}$  & {\bf N}  & $\Delta t$ & {\bf CFL} \\
\hline
 0.35 & 1.0 & 0.27852 & $6.1249747\times 10^{-3}$ & 8& $512^2$  & $10^{-4}$ & 0.2\\
\hline \hline
\end {tabular}
\label{table_cyl}
\vspace{8mm}
\end{table}
The problem was solved on a fixed $512^2$-point equi-spaced Cartesian grid, distributed on a $(2\pi)^2$ regular domain, using a fixed timestep size $\Delta t=2\times 10^{-4}{\rm Sec.}$, equivalent to ${\bf CFL} \approx 0.2$. The zero velocity margin $\delta=8\Delta x$ was chosen, which resulted in $504^2$ active grid points. On such a grid, there were about 28 grid points across the cylinder diameter, and the number of numerical boundary points was about 30 in average.\\
To keep the solution as smooth as possible, the solution was started from zero velocity at $\theta=-\frac{\pi}{2}$ phase angle, and therefore, a $0.25~ {\rm Sec.}$ time shifting is used when comparing our results with the other data.  Although the no-slip conditions are not satisfied accurately in the initial time steps (as discussed in $\S$ \ref{sec2.3}, and also see \cite{Wang, Uhlmann}), it developed appropriately in the next times, and therefore the results for ${\cal N}_r=1$ are presented here.
\begin{figure}[t]
\setlength{\unitlength}{1mm}
\includegraphics[width=0.48\textwidth]{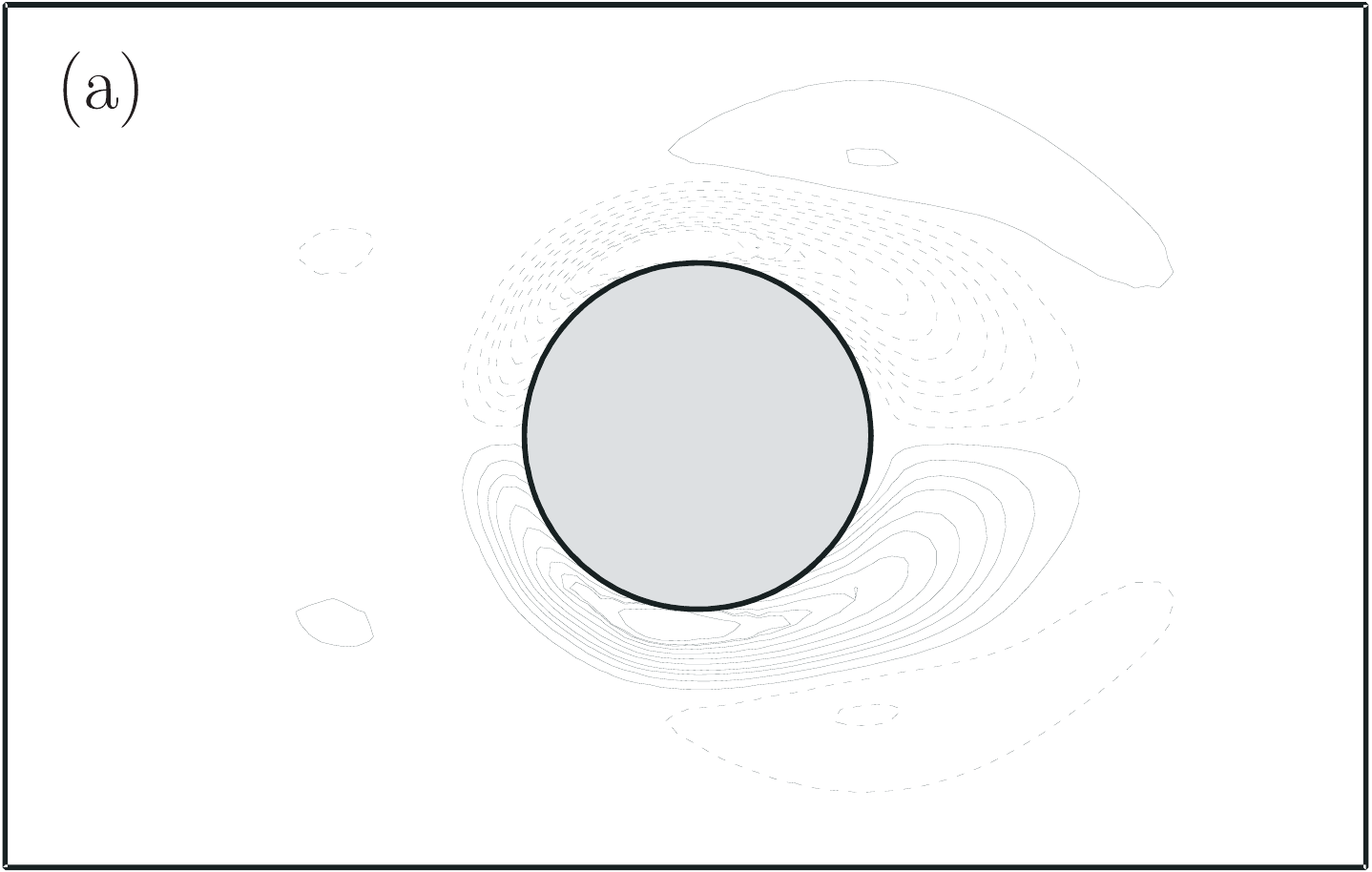}
\includegraphics[width=0.48\textwidth]{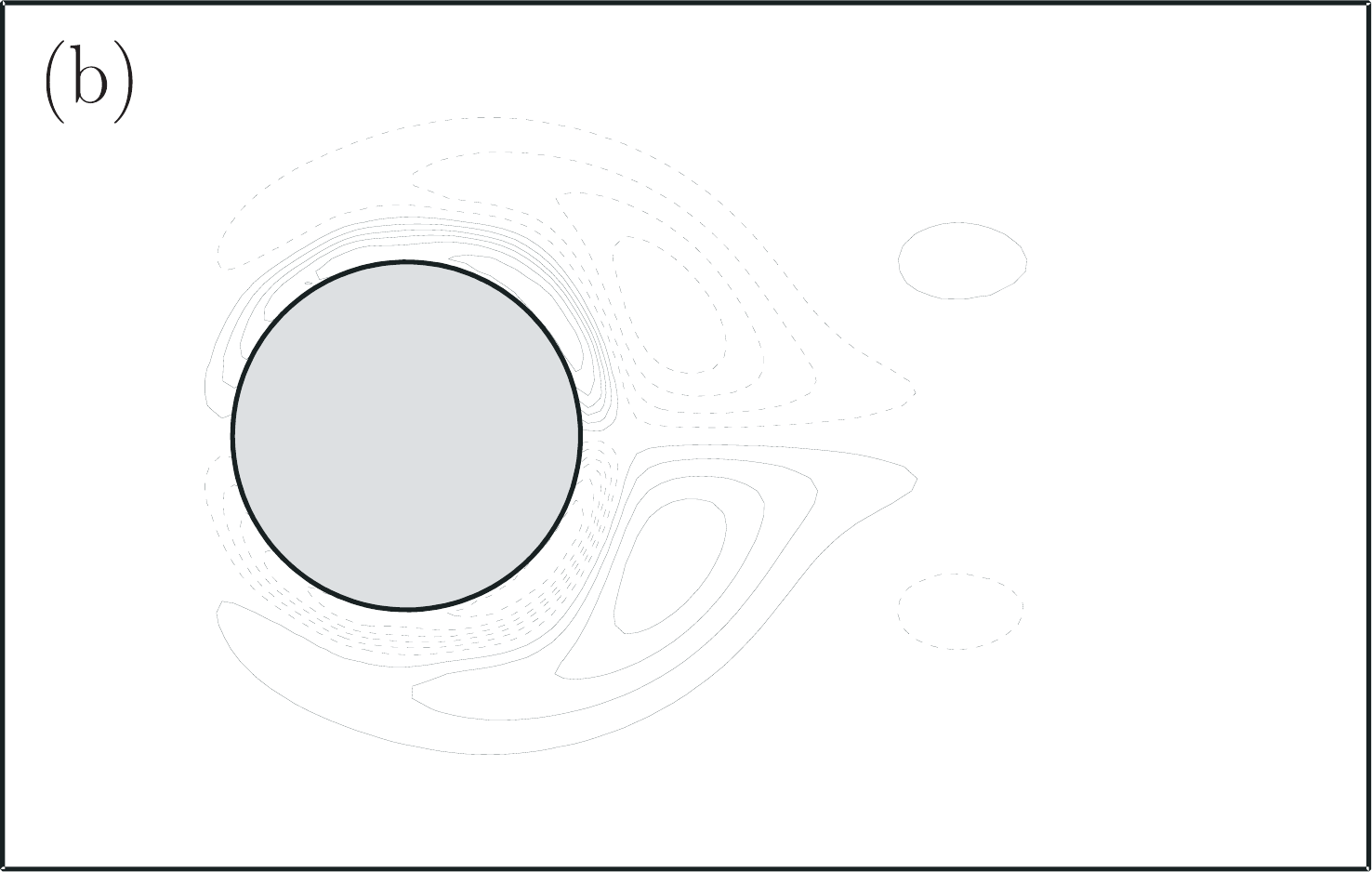} \\
\includegraphics[width=0.48\textwidth]{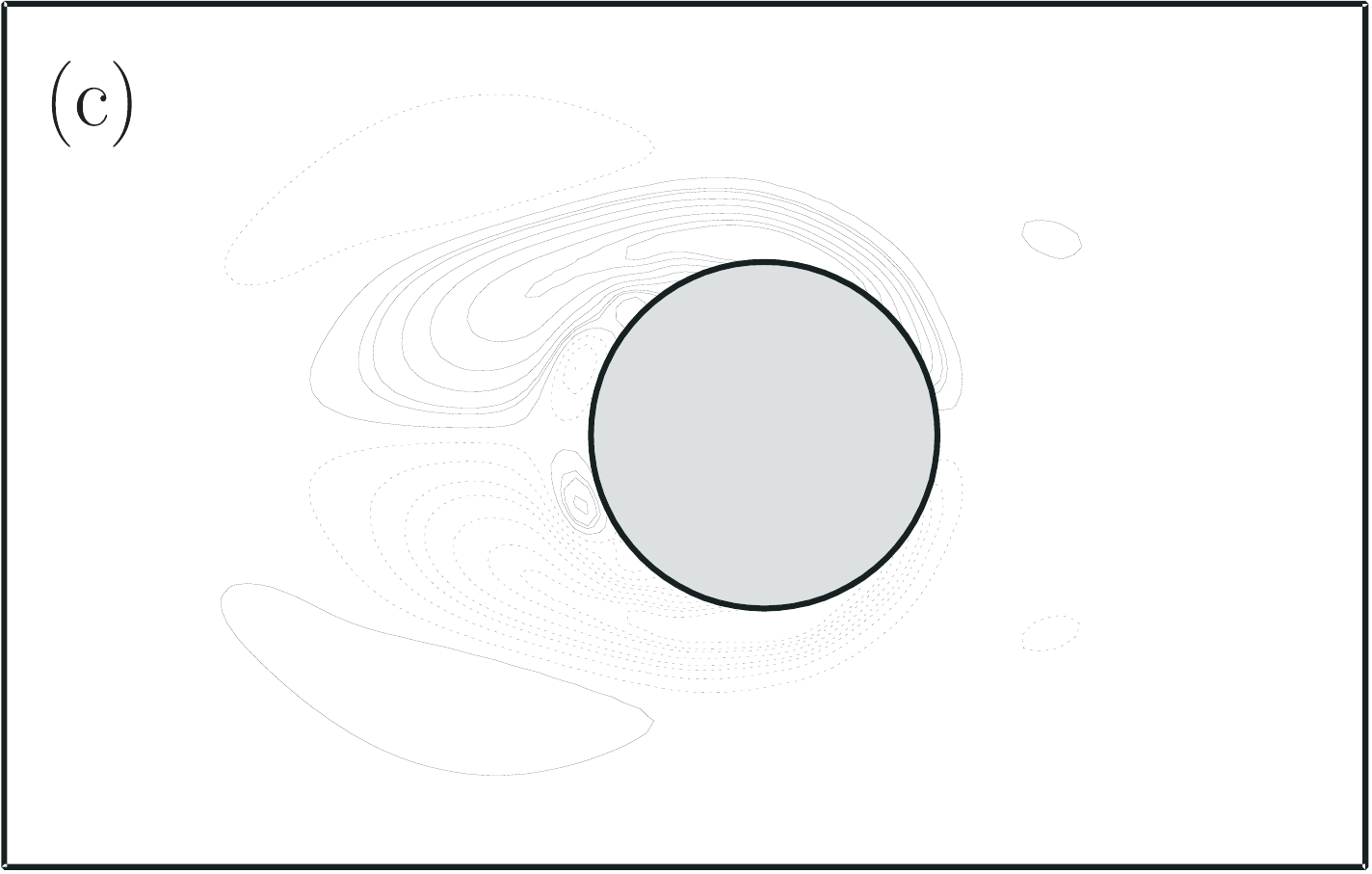}
\includegraphics[width=0.48\textwidth]{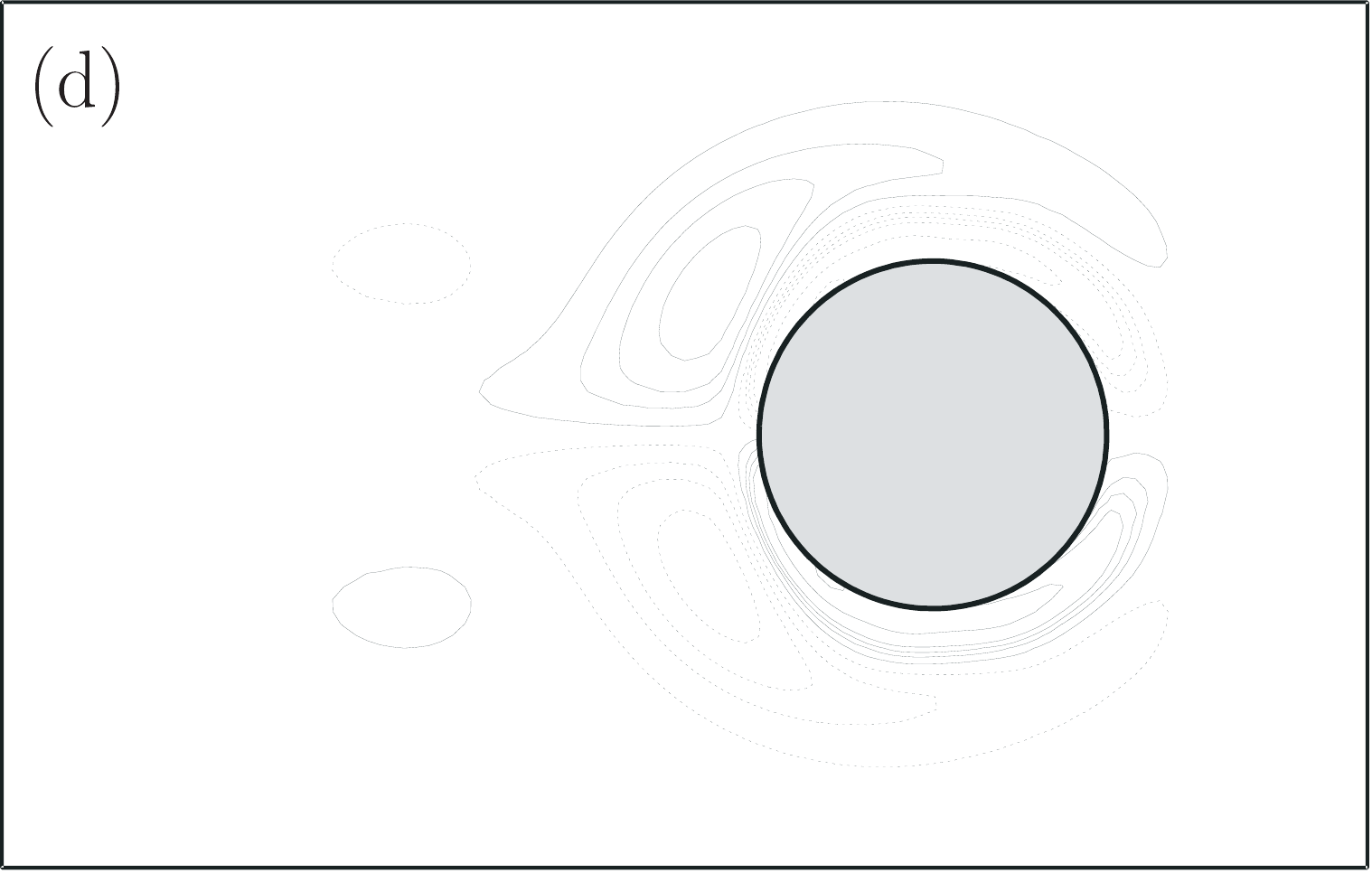}
\caption{Vorticity iso-lines of four phase angles {(a): $0^{\circ}$}, {(b): $96^{\circ}$}, {(c): $192^{\circ}$}, {(d): $288^{\circ}$}. The Helmholtz filter (with ${\bf C}_{\alpha}=0.46$) is used as a post-processor in order to remove the Gibbs oscillations  (see $\S$ \ref{Helm}). The dashed lines denote negative values. } \label{fig2cyl}
\vspace{5mm}
\end{figure}
\subsubsection{The captured dynamics}
The problem was solved for three different orders of boundary conditions ({\it i.e.}, ${\cal N}_p=0,1,2$), and the effects of ${\cal N}_p$ will be discussed later. Before that, the captured dynamics for ${\cal N}_p=2$ is presented here.\\
\begin{figure}[t]
\setlength{\unitlength}{1mm}
\centerline{\includegraphics[width=11cm]{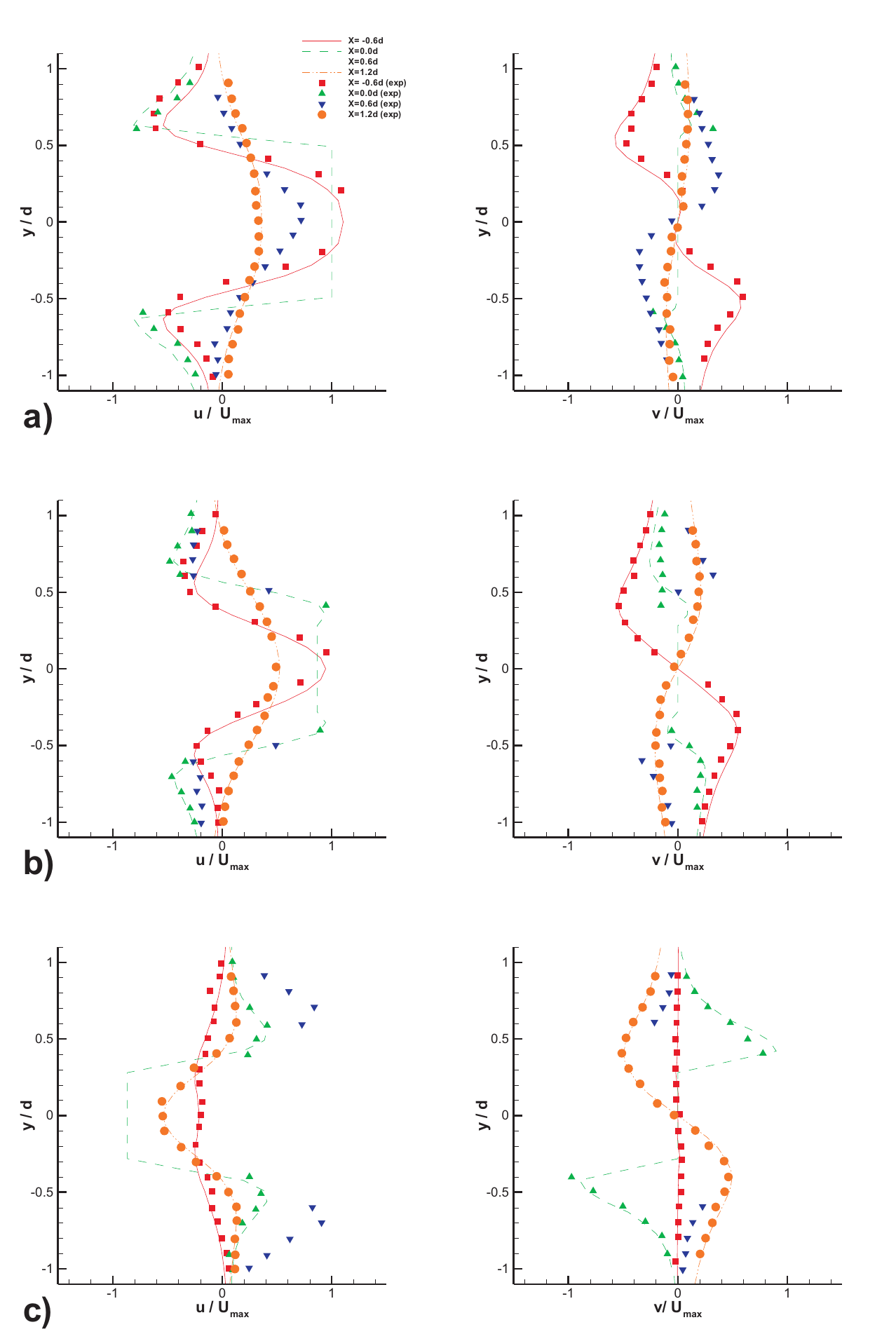}}
\caption{Comparison of the instantanious velocity profiles for three phase angles (a): $180^{\circ}$ , (b): $210^{\circ}$ , and (c): $330^{\circ}$. Lines show the results of the present method (obtained from the second-order boundary conditions ${\cal N}_p=2$), while the symbols are the experimental data  \cite{Dutsch}.}
\vspace{5mm}
\label{fig3cyl}
\end{figure}
The obtained vorticity dynamics is presented in  Fig. \ref{fig2cyl}. As one can see, it is in good agreement with the reported experimental and numerical studies of this $({\bf Re},{\bf KC})$ combination (c.f. \cite{Dutsch,Balaras,McDonough}). As the cylinder moves to one side, upper and lower boundary layers develop with symmetric starting and separation points; while a wake develops in the downstream of the cylinder, containing two symmetric counter-rotating vortices. Then in the opposite side cycle similar structures are observable, and the last wake vortices (now located in  front of the cylinder), split by the cylinder. More quantitative comparisons can be made by comparing the instantaneous velocity profiles, as illustrated in Fig. \ref{fig3cyl}. The velocities are compared with the experimental data of Dautch {\it et al.} \cite{Dutsch}, and similar to some other works \cite{Dutsch,Balaras,McDonough}, four $x_1/D$ sections are illustrated for each of the three phase angles. As one can see, there is a general agreement between the present results and the experimental data.\\
It should be noted that because of high levels of oscillations (especially for the vorticity fields), the aforementioned Helmholtz filter (see $\S$ \ref{Helm}) with ${\bf C}_{\alpha}=0.46$ was used (as a post-processor), in obtaining the vorticities of Fig. \ref{fig2cyl}. The ${\bf C}_{\alpha}$ was chosen by observation, such that the active contour levels can be compared with the benchmark data. Obtaining fairly acceptable results from an oscillating solution by use of a simple filter (just as a post-processor) is interesting, and it may be interpreted as a confirmation of the previous observations ({\it e.g.}, \cite{Gotlib, Keetels, Kolomenskiy}) that the higher degrees of pointwise accuracies can be recovered from a contaminated spectral solution (at least in some circumstances).\\
\begin{figure}[t]
\setlength{\unitlength}{1mm}
\centerline{\includegraphics[width=9cm]{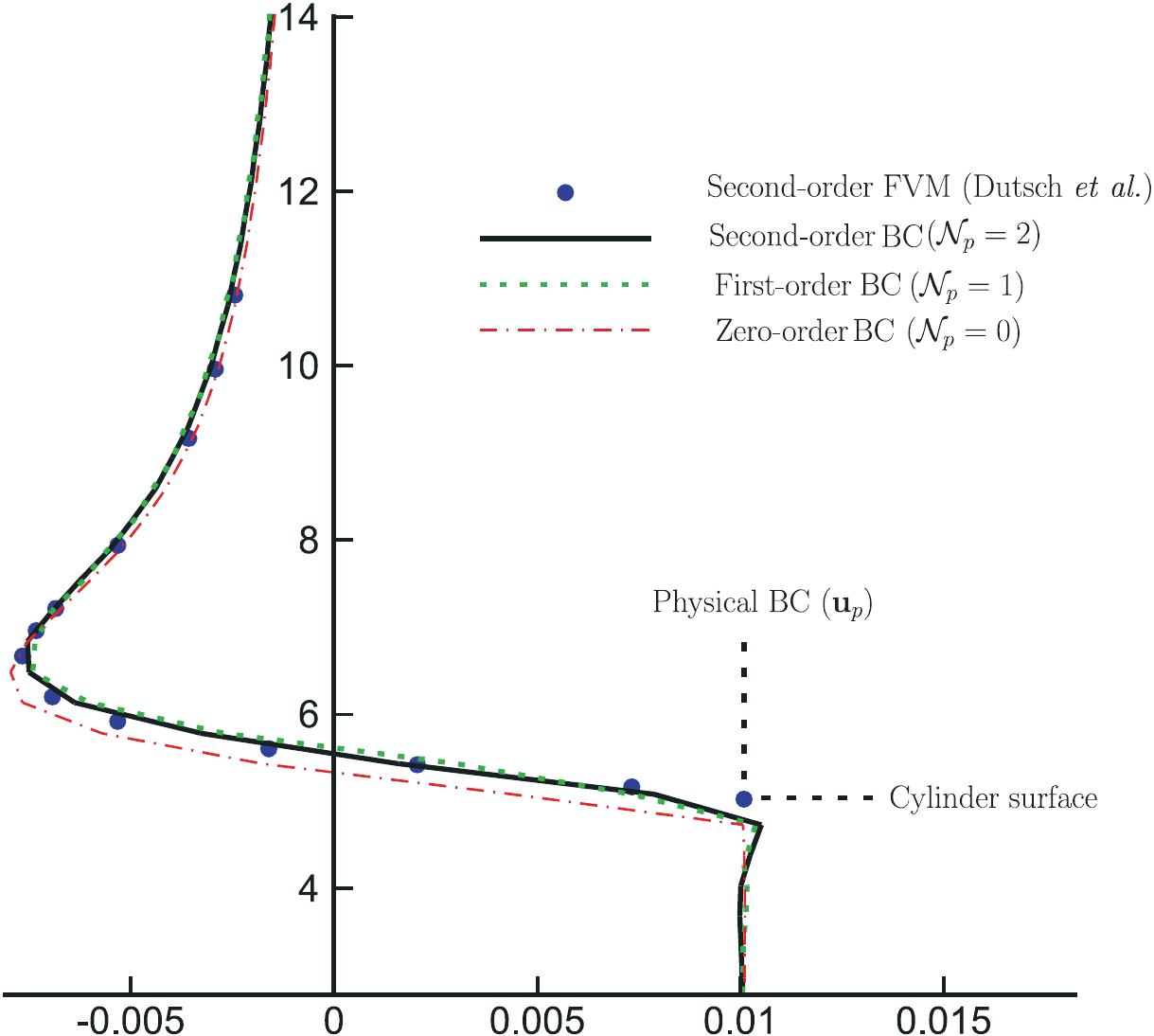}}
\caption{Longitudinal velocity profiles at $x_1/D=0$ of the $0^{\circ}$ phase angle. The results of different ${\cal N}_p$ are compared with the second-order finite volume solution of Dutsch {\it et al.} \cite{Dutsch}.}
\vspace{5mm}
\label{order_bc}
\end{figure}
\subsubsection{On the effects of order of the boundary condition setting}
Presence of various numerical and experimental studies of this problem offers a unique opportunity for direct evaluation of the effects of order of implementation of the boundary conditions ({\it i.e.}, ${\cal N}_p$), on the overall accuracy of the solution. In the last test case ({\it i.e.}, the dipole--wall collision), the ${\cal L}^p$ convergence of the vorticity fields were considered, which were highly under the influences of the Gibbs oscillations (see $\S$\ref{conv_rate}, and Figs. \ref{fig12} and \ref{fig13}). Instead, in the present section, it is aimed to evaluate the accuracy of the solution by direct comparison of the velocities with a second-order solution, which has been verified several times (that is, the Dutsch {\it et al.} solution  \cite{Dutsch}).\\
In Fig. \ref{order_bc} the longitudinal velocity for different orders of implementation of the boundary conditions (that is, ${\cal N}_p$=0,1,2), are compared with the second-order finite volume solution of Dutsch {\it et al.} \cite{Dutsch} in the $x/D=0$ section of the $0^{\circ}$ phase angle. This position is chosen intentionally because the effects of accuracy of the longitudinal velocity boundary condition can be evaluated directly. The following points can be made with regard to this figure:
\begin{enumerate}
\item[{\bf i)}] As one can see, the first-order and second-order implementation of the boundary conditions ${\cal N}_p=1,2$, yielded velocity profiles that are very close to the second-order finite volume solution. Therefore, we can say that for these orders of boundary conditions, the solution has been practically equivalent to a second-order solution. Comparisons of the velocities in the other phase angles and other $x/D$ sections (are not presented here for the sake of brevity), do confirm the above statement.
\item[{\bf ii)}] Even for the least order of boundary conditions (that is ${\cal N}_p=0$), except for the regions near to the cylinder surface, the velocities are again really close to the second-order finite volume solutions. This issue is practically important, since it justifies use of easy-to-implement case ${\cal N}_p=0$ (which bypasses some tedious geometric operations), in rough solutions for even very complex geometries.
\item[{\bf iii)}] Our numerical experiments have shown that the minimum amounts of the Gibbs oscillations occurs for ${\cal N}_p=0$, and the maximum occurs for ${\cal N}_p=2$. Note that the velocities are presented here without any kind of filtering. In fact, just for presentation of the vorticity fields (in Fig. \ref{fig2cyl}) the Helmholtz filter is used, because of high levels of the Gibbs oscillations.\\
\end{enumerate}
\begin{figure}[t]
\setlength{\unitlength}{1mm}
\centerline{\includegraphics[width=9cm]{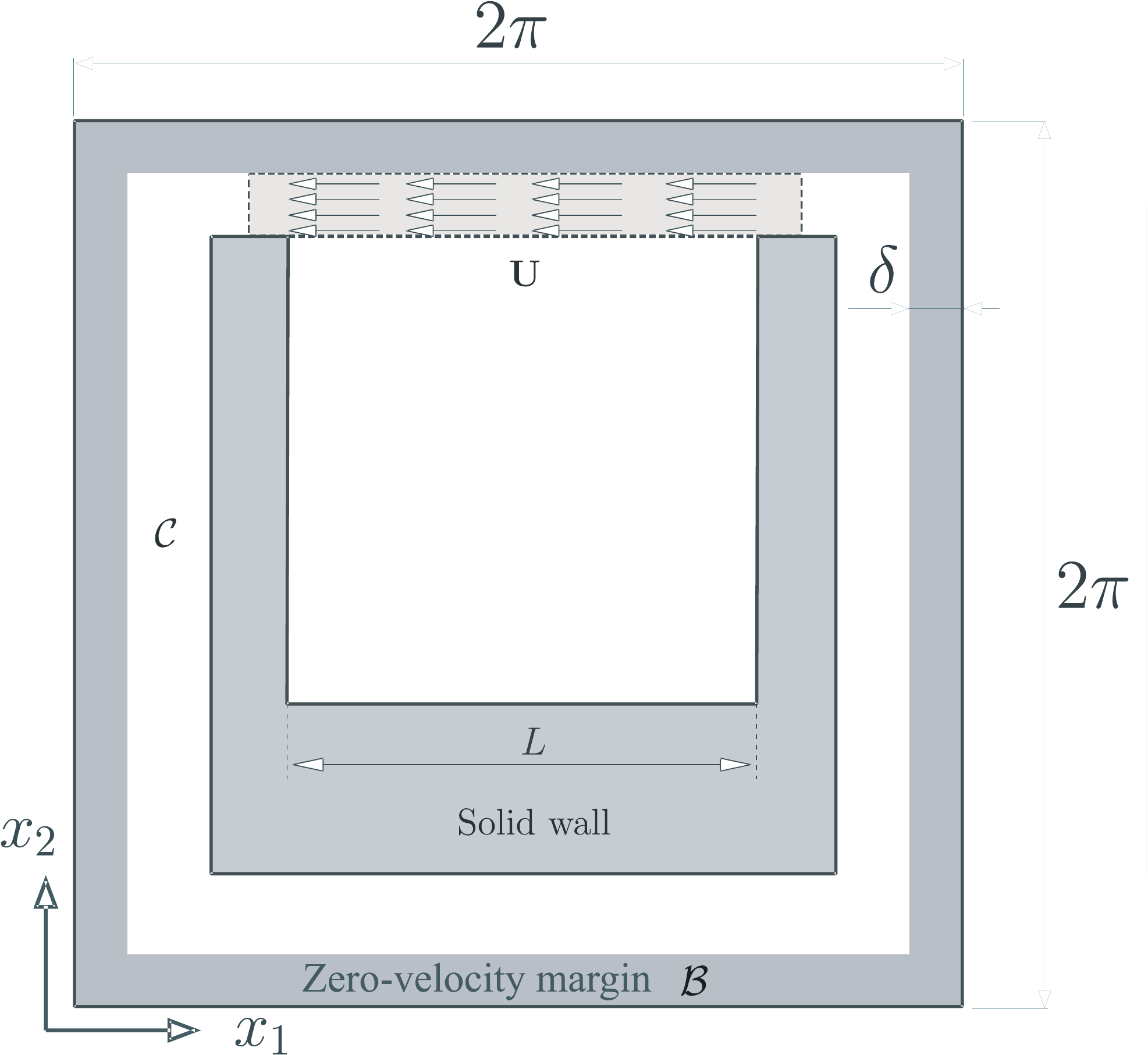}}
\caption{Geometry of the lid-driven cavity problem. The side chanel $\cal C$ is considered in order to balance the mass flow rate and acheiving zero mean velocity and vorticity.}
\label{fig1cav}
\vspace{5mm}
\end{figure}
\subsection{Lid-driven cavity flow}
The method is not restricted to the no-slip and no-penetration immersed boundary conditions. In fact,  any arbitrary immersed velocity boundary condition (which can be presented by a window or mask function), can be implemented via the proposed algorithm. To show this capability of the method, the classical two-dimensional lid-driven cavity flow is investigated in two different regimes. For the steady solutions, a fairly low Reynolds number flow {\bf Re}=100 is chosen, while the unsteady solutions is examined by analysis of a higher Reynolds number flow, that is {\bf Re}=1000.  Table \ref{cavitytable} presents the physical and numerical characteristics of the solutions.\\
Our suggested configuration is illustrated in Fig. \ref{fig1cav}. As one can see, the solid walls of cavity are implemented by a U-shaped solid immersed body, and a side channel (denoted by $\cal C$), is considered in order to balance the overall mass flow rate, and achieving zero mean velocity and vorticity fields. Both no-slip conditions and given velocity {\bf U} are set at once in the beginning of each time step, as explained in $\S$ \ref{sec2.3}. Since the cavity walls were coinciding with the Cartesian grid, the numerical and physical boundary points are coinciding, and therefore, extrapolation process is bypassed (similar to the dipole--solid wall problem).\\
\begin{table}[b]
\caption{Physical and numerical parameters of the cavity flow.} \label{tab4} \vspace{2mm}
\begin{tabular}{lccccc} \hline \hline
   &   {\bf Re}        & Grid resolution &  Active grid resolution  & $\Delta t$ & {\bf CFL} \\ \hline
steady       & 100    &     $256^2$         &      $185^2$           & $2\times 10^{-3}$     & $\approx 0.5$  \\
unsteady      & 1000    &     $512^2$         &     $375^2$            & $10^{-4}$    & $\approx 0.5$   \\ \hline \hline
\\
\end{tabular}
\label{cavitytable}
\end{table}
At first we examine the steady solution. Several numerical and experimental studies have demonstrated presence of steady solution for ${\bf Re}={\rm {\bf U}} L/\nu=100$ (see {\it e.g.} \cite{Ghia,Auteri1,Auteri2,Migeon}). In the present work, the problem was solved on a $256^2$-point grid, which because of presence of the margin $\cal B$, the side channel $\cal C$, and the solid U-shaped body, there were $185^2$ active grid points inside the cavity. Since the computational cost of the method scales by $(N\log N)$, one can said that about $48\%$ extra cost was paid for a $185^2$-point simulation. The solution begun from zero velocity field, and a constant time step $\Delta t=2\times 10^{-3}$ Sec. (equivalent to ${\bf CFL}\approx 0.5$) was used; and the solution was continued until the steady solution was established after about $8.6$ Sec.\\
\begin{figure}[t]
\setlength{\unitlength}{1mm}
\includegraphics[width=0.46\textwidth]{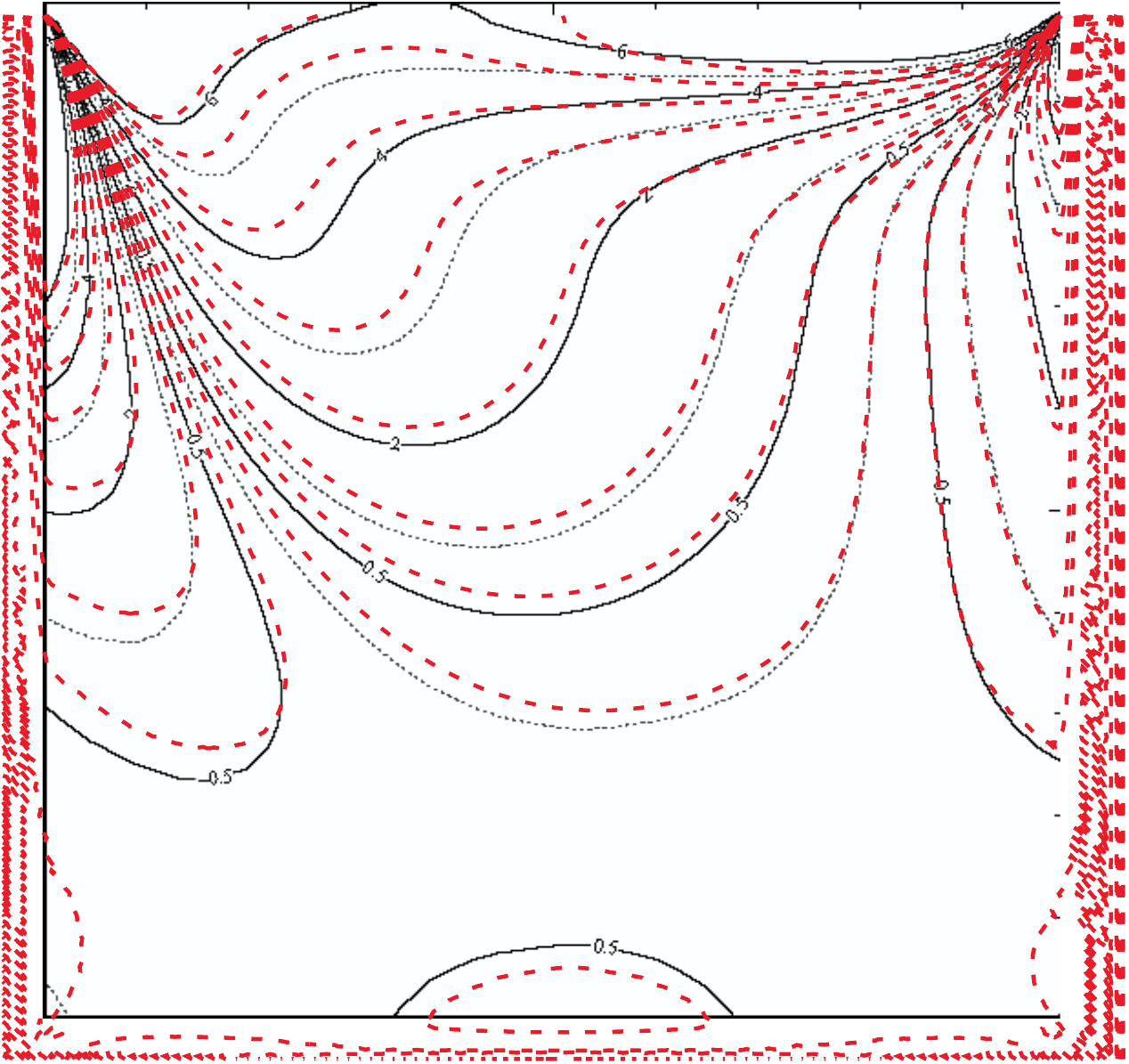}
\includegraphics[width=0.49\textwidth]{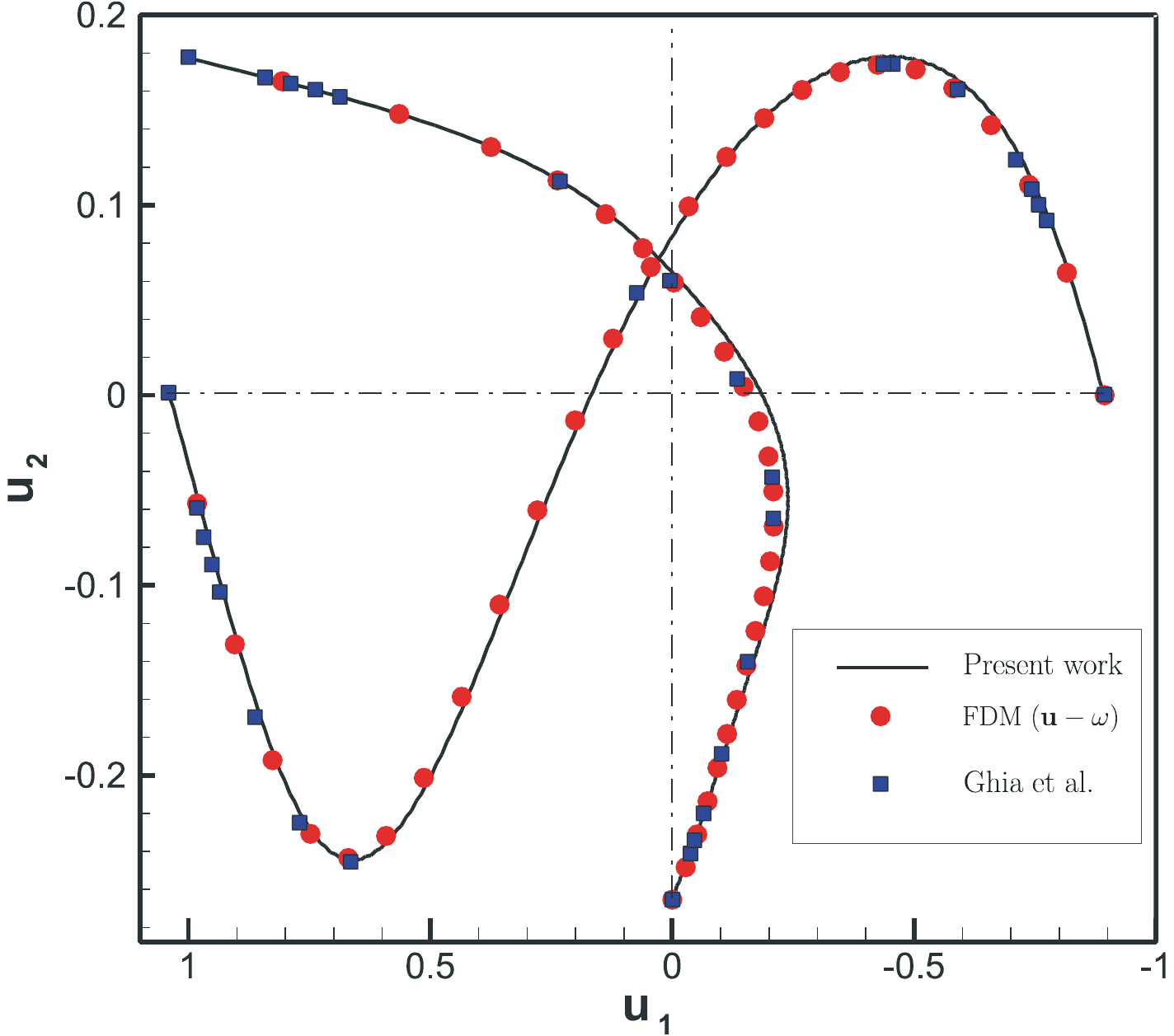}
\caption{{\bf Left:} The vorticity iso-lines of the steady solution ({\bf Re}=100) in comparison to the results of Auteri {\it et al.} \cite{Auteri1}. The dashed lines show the results of the present work. {\bf Right:} The velocity profiles in comparison to the results of Ghia {\it et al.} \cite{Ghia} and a classical second-order finite difference vorticity--velocity solution.}
\label{fig2cav}
\vspace{5mm}
\end{figure}
On the other hand, the numerical experiments showed that ${\cal N}_r>1$ was needed for adequate satisfaction of the no-slip conditions. It can be a consequence of dominance of the solid walls on the solution in this particular flow. In fact, when implementing the boundary conditions, a huge amount of discontinuities are imposed to the ${\bf u}^{\rm BC}$ (in comparison to {\it e.g.}, the oscillating cylinder problem). Therefore, removing these discontinuities, and enforcing the solenoidality, changes the immersed boundary conditions of ${\bf u}^{\rm BC}_{\bf Sol}$ such that it is needed to set them more than once (see $\S$ \ref{sec2.3}). However, our experiments did not show changes in the results for ${\cal N}_r>3$, and therefore, the present results are obtained with ${\cal N}_r=3$. \\
The results are shown in Fig. \ref{fig2cav}. In the left panel, the vorticity iso-lines are compared with the results of singularity-removed Chebyshev solution of Auteri {\it et al.} \cite{Auteri1}. As one can see, there is a good overall agreement between the results, although some discrepancies (especially in the middle of the cavity) are noticeable. However it should be emphasized that there are substantial differences between our solution and the Auteri {\it et al.} method \cite{Auteri1}, in implementation of the boundary conditions and levels of solenoidality. In the right side of Fig. \ref{fig2cav}, the velocity profiles in the middle of cavity are presented and compared with the data of Ghia {\it et al.} \cite{Ghia}, and a classical second-order finite difference solution of vorticity--velocity formulation of the NSE. In the finite difference solution, the no-slip conditions are implemented via the explicit Thom's rule \cite{Russel}, which satisfies just approximately the solenoidality \cite{Rempfer}. As it can be seen, while the Ghia {\it et al.} data (which obtained from solution of the primitive variables formulation of the NSE, and pure Neumann boundary conditions for the Poisson pressure equation), are coinciding with the finite difference results; they show some discrepancies with our data, especially far from the solid walls.\\
\begin{figure}[t]
\setlength{\unitlength}{1mm}
\centerline{\includegraphics[width=8cm]{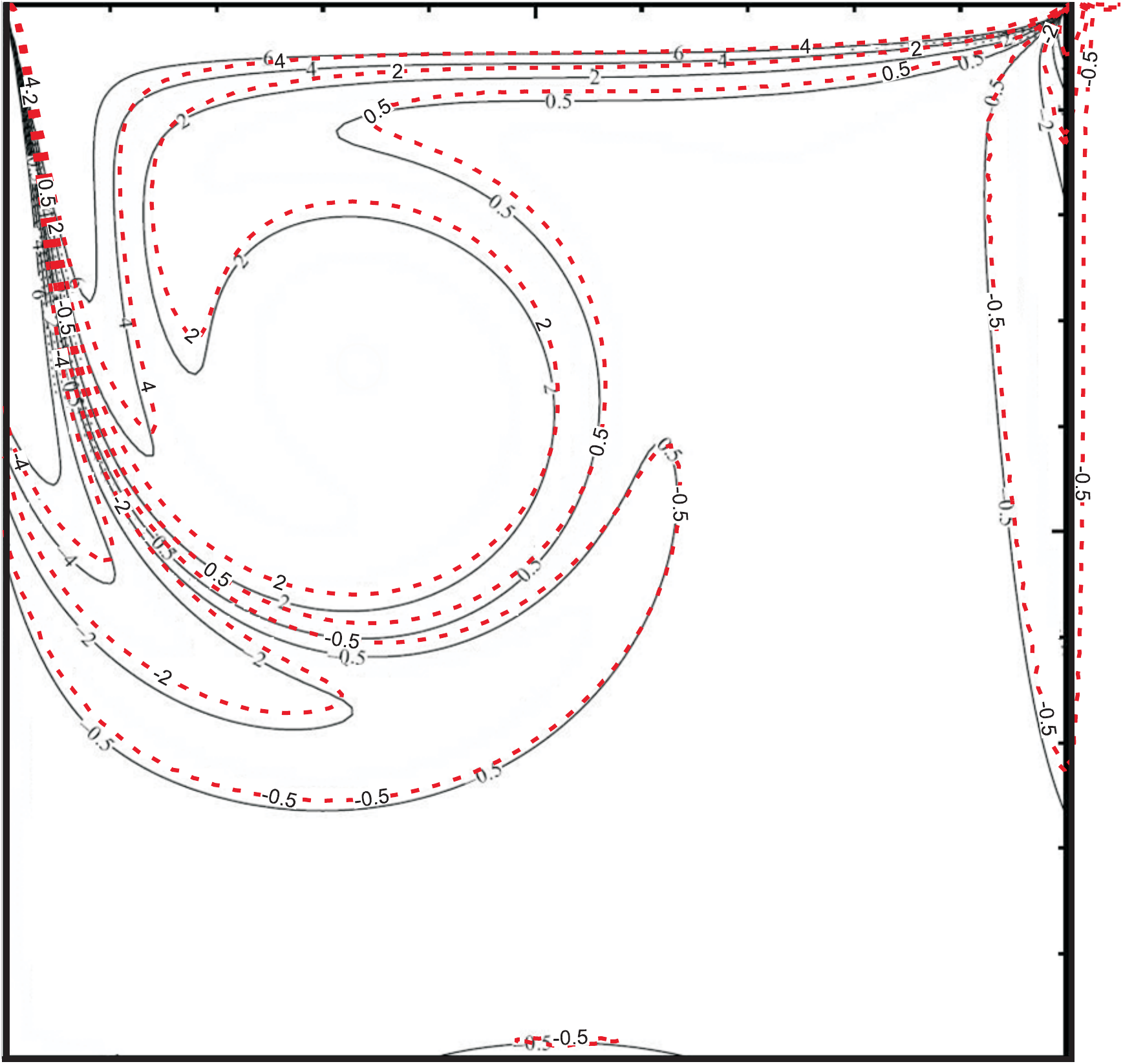}}
\caption{The vorticity iso-lines of the unsteady solution for {\bf Re}=1000, at the time instant $t=6.25$ Sec, compared with the data of Auteri {\it et al.} \cite{Auteri1}. The dashed lines show the results of the present method.}
\label{fig3cav}
\vspace{5mm}
\end{figure}
In addition to the steady solution, it is aimed to verify time accuracy of the method. To this end, a higher Reynolds number (that is, {\bf Re}=1000) in an unsteady regime is examined. The unsteady cavity flow has studied several times experimentally and numerically (see {\it e.g.} \cite{Auteri1,Auteri2,Migeon}). The problem was solved on a $512^2$-point equi-spaced Cartesian grid which resulted in $375^2$ active grid points inside of the cavity. Therefore, the extra cost of the calculations has been about $46 \%$. To keep the {\bf CFL} number about 0.5, a constant time step $\Delta t=10^{-4}$ was chosen. The solution was begun from a zero velocity field at $t=0$, and continued until $t=6.25$, which the results are comparable by the other results \cite{Auteri1, Auteri2}. Like the last test case, ${\cal N}_r=3$ was used. In Fig. \ref{fig3cav} the vorticity iso-lines in  comparison to the results of singularity-removed Chebyshev solution \cite{Auteri1} are shown. Good agreements between the results are observable, despite the fact that a different time integration method (that is, the second-order Adams--Bashforh method) is used in obtaining the results of \cite{Auteri1} and \cite{Auteri2}.
\section{Conclusions}
An immersed boundary Fourier pseudo-spectral method was proposed for the vorticity-velocity formulation of the two-dimensional incompressible NSE. The zero-mean Fourier pseudo-spectral solution are used, and therefore, the method is applicable to the confined flows, which are modeled by considering a zero velocity margin in the vicinity of the regular boundaries. Without explicit addition of external forcing functions, arbitrary Dirichlet velocity boundary conditions are implemented by direct modification of the diffusion and convection terms of the vorticity transport equation; and in this way, it was shown that the obtained velocities are solenoidal. The immersed boundary conditions are approximated on some regular grid points (called the numerical boundary points), by use of different orders (up to second-order) polynomial extrapolations, in the normal directions to the immersed surfaces.\\
Although because of presence of the Gibbs phenomenon the ${\cal L}^p$ spatial rates of convergence are not so high (even less than two, especially in the regions near the solid walls), good agreements between our results with some other second-order solutions were observed in some test cases. On the other hand, the temporal rate of convergence was found to be a function of timestep sizes, and for sufficiently small timestep sizes, the highest order of time discretization (that is, the fourth order) was obtained.\\
Like many other immersed boundary methods, it was observed that implementation of the immersed velocity boundary conditions imposes some discontinuities; and conversely, imposing the continuity changes the immersed boundary conditions. However, it was observed that by repeating the procedure, and by development of the solution, some solenoidal velocities, which satisfy the immersed boundary conditions (approximately) are obtainable.\\
The method was implemented to a conventional Fourier pseudo-spectral solver of the vorticity-velocity form of the NSE, without any changes in the internal structure of the solver, and the performance of the method was compared with the original solver. In this context, it was observed that fixed and moving immersed boundaries can be treated by a computational cost, which scales by ($N\log N$).\\
The best accuracies (and also the most Gibbs oscillations), were obtained for the second-order boundary conditions. However, even for the easy-to-implement zero-order boundary conditions, fairly accurate solutions were obtained, at least far from the solid walls. Particularly this feature of the method is interesting, because this simple algorithm can be used in a rather wide range of applications, from rough simulations to the fairly accurate ones, just by changing the order of implementation of the boundary conditions. In our opinion, it is particularly a consequence of using a pseudo-spectral solver as the core of the method.\\
Just a simple constant-width Helmholtz filter was used in the present work,  as a postprocessor, and in order to remove the Gibbs oscillations in visualization of the results. However, like other spectral methods, one can use some particular numerical filters to improve the rates of convergence. This is one of several lines which we can assume for the extension of the present method.
\newpage
\appendix{\bf{Appendix: Some fundamental considerations about the proposed algorithm}\\} \label{appendix}
Since the zero-mean Fourier series are used in the present method, the method is applicable to the flows with zero-mean velocity and vorticity, and with zero dynamics of the mean vorticity (see Eqns. (\ref{e3}) or (\ref{e3p})). Now from the Stokes theorem, the mean vorticity reads
\begin{equation}
\hat{\omega}(0,0)=\bar{\omega}=\frac{1}{\Omega_D}\oint_{\Gamma_D}{\bf {\rm } u}(\Gamma_D) \cdot d\Gamma_D.
\label{Stokes1}
\end{equation}
Among many flow configurations with $\bar{\omega}=0$, the ones that ${\bf {\rm u}}_{\Gamma_D}=0$ are dealt in the present work, and these flows are modeled by considering a zero-velocity margin, as it is shown in Fig.~\ref{fig1}. \\
With this in mind, the present appendix is devoted to explanation of three fundamental issues, particularly related to steps 3 and 4 of the proposed algorithm ($\S$ \ref{sec2.3}). These discussions are presented here separately for the sake of clarity, and to prevent disturbing the main flow of  the paper.
\subsection{The mean value of the conditioned vorticity}
Clearly, Eq. (\ref{e7}) yields a zero-mean conditioned vorticity (since $k_1=k_2=0$ results in $\hat{\omega}^{\rm BC}(0,0)=0$), as it is desired. However, legitimacy of application of this equation for arbitrary ${\bf {\rm u}}^{BC}$ (obtained from arbitrary immersed boundary conditions and different window functions), needs some considerations.\\
In general, the Stokes theorem states that for a confined flow (${\bf u}_{\Gamma_D}=0$) the mean vorticity is vanished, regardless of distribution of velocities in $D$ (see Eq. \ref{Stokes1}). However, the point is that the Stokes theorem is applicable to the velocity fields that are $C^1$ at least; the condition which can be violated in the general immersed boundary condition settings and windowings.\\
On the other hand, note that it is desired to find the solution in the Fourier space, and regardless of rate of decaying of $\hat{\bf u}^{\rm BC}={\rm {\bf FT}} \{ {\bf u}^{\rm BC} \}$ (which is depended on the smoothness of ${\bf u}^{\rm BC}$), however, ${\rm {\bf FT}}^{-1} \{ \hat{\bf u}^{\rm BC} \}$ has $C^{\infty}$ smoothness (since it is constructed from the Fourier functions which are $C^{\infty}$ smooth). Now, by applying the Stokes theorem on ${\rm {\bf FT}}^{-1} \{ {\rm {\bf FT}} \{ {\bf u}^{\rm BC} \} \}$, one can find
\begin{equation}
\bar{\omega}^{\rm BC}=\frac{1}{\Omega_D}\int_{\Omega_D} (\nabla \times {\bf u}^{\rm BC}) d\Omega_D =\frac{1}{\Omega_D} \oint_{\Gamma_D} {\rm {\bf FT}}^{-1} \{ {\rm {\bf FT}} \{ {\bf u}^{\rm BC} \} \}(\Gamma_D)d{\Gamma_D}.
\label{Stokes}
\end{equation}
On the other hand, it should be noted that ${\rm {\bf FT}} \{ {\bf u}^{\rm BC} \}$ can be suffered from the Gibbs oscillations because of presence of discontinuities in ${\bf u}^{\rm BC}$, and therefore, ${\rm {\bf FT}}^{-1} \{ {\rm {\bf FT}} \{ {\bf u}^{\rm BC} \} \}$ is not necessarily zero on ${\Gamma_D}$. Consequently, $\bar{\omega}^{\rm BC}$ is not vanishing in general.\\
However, since the zero-mean flows are dealing, these $\bar{\omega}^{\rm BC}$ supposed to be discrepancies from the flow physics, and they are ignored in the solution procedure. On the other hand, as our numerical experiments were shown, these spurious mean vorticities decrease as the solution is developed; and for the logical timestep sizes, they get practically-ignorable values almost immediately after beginning of the solution.
\subsection{On the solenoidality of the velocities}
A key feature of the present method is that the physical (solenoidal) velocities are obtained and used, regardless of complexity of the immersed boundary conditions. Here it is desired to justify why Eq. (\ref{e5p}) results in solenoidal velocities for an arbitrary $\omega^{\rm BC}$.\\
\begin{prop}\label{prop2}
For any conditioned vorticity $\omega^{\rm BC}\in {\cal L}^2(\bar{D})$ the velocity vector
\begin{equation}
\hat{\bf u}_{\bf Sol}^{\rm BC}=-i\frac{{\bf k}^{\bot}}{|{\bf k}|^2}\hat{\omega}^{\rm BC}
\label{app1}
\end{equation}
is solenoidal, in which $\hat{\omega}^{\rm BC}={\rm {\bf FT}} \{ {\omega}^{\rm BC}\}$.\\
\end{prop}
Note that Eq. (\ref{e5p}) is re-written again (that is Eq. \ref{app1}) for clarity.
\begin{pf}
There are many ways to proof the above proposition, and we give one of them which is more consistent with our purposes.\\
At first, since $\omega^{\rm BC}\in {\cal L}^2(\bar{D})$, the Fourier coefficients $\hat{\omega}^{\rm BC}({\bf k})={\rm {\bf FT}} \{ {\omega}^{\rm BC}\}$ exist for any $\bf k$, although their rate of decaying can be not so high. Now, it is a well-known fact  \cite{Canuto1} that the continuity equation in the Fourier space reduces to
\begin{equation}
{\bf k}\cdot {\hat {\bf u}}_{\rm {\bf Sol}}=0.
\end{equation}
In the other words, the solenoidal velocities are perpendicular to the wavenumber vector and vice versa. Now referring to Eq. (\ref{app1}), the proposition is proven since the velocities $\hat{\bf u}_{\bf Sol}^{\rm BC}$ are obtained from ${\bf k}^{\bot}$, and therefore they are perpendicular to ${\rm {\bf k}}$.\\
\rightline{$\blacksquare$}
\end{pf}
In addition to the above proof, also our extensive numerical experiments on different test cases have shown that the divergence of ${\bf u}_{\rm {\bf Sol}}^{{\rm BC}}$ remains in the order of machine accuracy everywhere in $\bar D$.\\
The above simple proof in the Fourier space should not hide the detailed mechanism that discards the divergence. In fact, by taking the divergence of both sides of Eq. (\ref{e2}) we have
\begin{equation}
\nabla^{2}(\nabla \cdot {\bf u})=\nabla \cdot ({\hat {\bf e}}_{z}\times\nabla\omega)=0, \label{divapp}
\end{equation}
which means the obtained velocities from Poisson equation Eq. (\ref{e2}) are such that the divergence of them is a harmonic function (and therefore gets its maximum values on the boundary $\Gamma_D$), regardless of the boundary conditions. As an immediate consequence, the divergence of the velocities can be controlled by controlling the divergence on (near) the boundary $\Gamma_D$. This statement is not new, and it has been mentioned and discussed previously in many references (see {\it e.g.} \cite{Dennis} or \cite{Rempfer}).\\
Now definition of a zero velocity margin $\cal B$ in the vicinity of $\Gamma_D$ can be seen as a way for controlling the divergence. This strategy is not restricted to the spectral method, and can be used in other numerical methods, {\it e.g.} the finite difference method. However, in the Fourier pseudo-spectral solutions, since there is not any boundaries, the divergence vanishes perfectly everywhere on $D$, as it proved. In these cases the margin $\cal B$ helps achieving the higher rates of decaying of the Fourier coefficients.\\
In fact, we found that considering a closed flat zero velocity margin in the vicinity of the regular boundary (when it is possible), is a good idea for overcoming many difficulties related to finding appropriate vorticity boundary conditions and obtaining the solenoidal velocities. We will follow this idea in our next works on the fluid--solid interaction problems.
\subsection{About the dynamics of the mean vorticity}
There is still another question that should be answered with regard to the proposed algorithm. In fact, by time integration of Eq. (\ref{e3p}), we assumed zero dynamics for the mean vorticity during the time interval $\Delta t$. Legitimacy of this assumption is in order. \\
Note that in modeling of the flow configuration of Fig. \ref{fig1}, the modified ${\rm{\bf L}}$, ${\rm{\bf N}}_1$, and ${\rm{\bf N}}_2$ are introduced into the vorticity transport equation, and then this modified equation is transformed into the Fourier space (yielded Eq. (\ref{e3p})). Therefore, in order to analysis of the dynamics of the mean vorticity of our numerical model, the mean mode of the modified vorticity transport equation should be analyzed.\\
%
{\bf Theorem 1 } {\it Given a conditioned vorticity $\omega^{\rm BC}=\nabla\times {\bf u}^{{\rm BC}}$, time integration of
 }
\begin{equation}
\partial_t \omega=\nu\nabla^2\omega^{\rm BC} -\frac{\partial^2}{\partial x_1\partial x_2}(u_2^2-u_1^2)_{\rm{\bf Sol}}^{\rm BC}+(\frac{\partial^2}{\partial x_2^2}-\frac{\partial^2}{\partial x_1^2})(u_1u_2)_{\rm{\bf Sol}}^{\rm BC},\label{e1prop2}
\end{equation}
{\it results in zero dynamics for the mean vorticity, in which ${\bf u}_{\rm {\bf Sol}}^{{\rm BC}}={\rm {\bf FT}}^{-1} \{ \hat{\bf u}_{\bf Sol}^{\rm BC}  \}$, and $\hat{\bf u}_{\bf Sol}^{\rm BC}$ is obtained from Eq. (\ref{e5p}).
 }\\

{\bf PROOF.}
To show $\partial_t \bar{\omega}=0$, one way is to show that the modified ${\rm{\bf L}}$, ${\rm{\bf N}}_1$, and ${\rm{\bf N}}_2$ have zero means. Now, at first, note that the diffusion term ${\rm{\bf L}}$ (that is, the Laplacian of vorticity) is essentially silent about the mean values, since it consists of the (second-order) derivatives in both directions. On the other hand, the mean values of the other two convection terms (that is, ${\rm{\bf N}}_1$, and ${\rm{\bf N}}_2$) can be found directly. For example
\begin{equation}
(\bar{{\rm {\bf N}}}_1)_{\rm {\bf Sol}}^{\rm BC}=\frac{1}{l_1 l_2}{\mathcal D}^{-1}_{x_1}{\mathcal D}^{-1}_{x_2}( {\rm {\bf N}}_1)_{\rm {\bf Sol}}^{\rm BC}=\frac{1}{l_1 l_2}{\mathcal D}^{-1}_{x_1}{\mathcal D}^{-1}_{x_2}[{\mathcal D}_{x_1}{\mathcal D}_{x_2}(u_2^2-u_1^2)_{\rm {\bf Sol}}^{\rm BC}] \label{mean1},
\end{equation}
where ${\mathcal D}_{x_i}$ stands for derivative with respect to $x_i$; while $ {\mathcal D}^{-1}_{x_i}=\int_{0}^{l_i} dx_i$ shows the inverse of derivative operator $\mathcal D_{x_i}$ (that is, integration on the regular domain $\bar{D}$).  Now, implementation of the integral operators and changing the order yields
\begin{eqnarray}
(\bar{{\rm {\bf N}}}_1)_{\rm {\bf Sol}}^{\rm BC}&&=\frac{1}{l_1 l_2}({\mathcal D}^{-1}_{x_1}{\mathcal D}_{x_1})[{\mathcal D}^{-1}_{x_2}{\mathcal D}_{x_2} (u_2^2-u_1^2)_{\rm {\bf Sol}}^{\rm BC}] \nonumber \\
&&=\frac{1}{l_1 l_2}({\mathcal D}^{-1}_{x_1}{\mathcal D}_{x_1})[(u_2^2-u_1^2)_{\rm {\bf Sol}}^{\rm BC}\mid_0^{l_2}] \nonumber \\
&&=\frac{1}{l_1 l_2}[(u_2^2-u_1^2)_{\rm {\bf Sol}}^{\rm BC}\mid_0^{l_2}]\mid_0^{l_1}. \label{mean}
\end{eqnarray}
On the other hand, since ${\bf u}_{\bf Sol}^{\rm BC}$ is double periodic on $\bar{D}$, one can write
\begin{eqnarray}
&&(u_2^2-u_1^2)_{\rm {\bf Sol}}^{\rm BC}(x_1=0,x_2)=(u_2^2-u_1^2)_{\rm {\bf Sol}}^{\rm BC}(x_1=l_1,x_2), \nonumber\\
&&(u_2^2-u_1^2)_{\rm {\bf Sol}}^{\rm BC}(x_1,x_2=0)=(u_2^2-u_1^2)_{\rm {\bf Sol}}^{\rm BC}(x_1,x_2=l_2).
\end{eqnarray}
Substitution of these relations into Eq. (\ref{mean}), yields $(\bar{{\rm {\bf N}}}_1)_{\rm {\bf Sol}}^{\rm BC}=0$.\\
Exactly the same procedure can be followed for the $(\bar{{{\rm {\bf N}}}}_2)_{\rm {\bf Sol}}^{\rm BC}$ which results in
\begin{equation}
(\bar{{ {\rm {\bf N}}}}_2)_{\rm {\bf Sol}}^{\rm BC}=0.
\end{equation}
\begin{flushright}
$\blacksquare$
\end{flushright}
\begin{rem}
\rm Note that changing the derivatives order from Eq. (\ref{mean1}) to Eq. (\ref{mean}) needs ${\bf u}_{\bf Sol}^{\rm BC}\in C^1(\bar{D})$ at least. On the other hand, as it was argued before, ${\bf u}_{\bf Sol}^{\rm BC}={\rm {\bf FT}}^{-1} \{ {\rm {\bf FT \{ -\nabla \times \omega^{\rm BC} \}}} \} \in C^{\infty}(\bar{D})$ regardless of the rate of convergence of $\hat{\bf{\rm u}}_{\bf{\rm Sol}}^{\rm BC}$.
\end{rem}
\begin{rem}
\rm Although the theorem and proof are offered for this particular form of the vorticity transport equation, that is Eq. (\ref{e1mod}); it is worth mentioning that they are valid and applicable to the conventional form of vorticity transport equation (\ref{e1}), but it is just needed applying the Gauss theorem once, on each convection term.
\end{rem}

\end{document}